\documentclass[journal]{IEEEtran}
\usepackage{cite}
\usepackage{color,soul}
\usepackage{gensymb}
\usepackage{dsfont}

\ifCLASSINFOpdf
  \usepackage[pdftex]{graphicx}
  \graphicspath{{../pdf/}{../jpeg/}}
  \DeclareGraphicsExtensions{.pdf,.jpeg,.png}
\else
  \usepackage[dvips]{graphicx}
  \graphicspath{{../eps/}}
  \DeclareGraphicsExtensions{.eps}
\fi
\usepackage[cmex10]{amsmath}

\usepackage{amssymb}
\DeclareMathOperator*{\argmaxA}{arg\,max}{}

\usepackage{euscript}

\usepackage{multirow}
\usepackage{algorithm}
\usepackage{algpseudocode}
\usepackage{diagbox}
\usepackage{physics}
\usepackage{tikz}
\usepackage{comment}

\ifCLASSOPTIONcompsoc
  \usepackage[caption=false,font=normalsize,labelfont=sf,textfont=sf]{subfig}
\else
  \usepackage[caption=false,font=footnotesize]{subfig}
\fi
\usepackage{url}

\begin{document}
%
\title{A Learning-based Power Management for Networked Microgrids Under Incomplete Information }
%
%
%

\author{
         Qianzhi Zhang,~\IEEEmembership{Student Member,~IEEE,}
         Kaveh Dehghanpour,~\IEEEmembership{Member,~IEEE,}
         Zhaoyu Wang,~\IEEEmembership{Member,~IEEE,} 
         
         and Qiuhua Huang,~\IEEEmembership{Member,~IEEE}
\thanks{
Q. Zhang, K. Dehghanpour, and Z. Wang are with the Department of
Electrical and Computer Engineering, Iowa State University, Ames,
IA 50011 USA (e-mail: wzy@iastate.edu).

Q. Huang is with Pacific Northwest National Laboratory, Richland, WA, 99354 USA (e-mail: qiuhua.huang@pnnl.gov).
}
}
\maketitle

\begin{abstract}
This paper presents an approximate Reinforcement Learning (RL) methodology for bi-level power management of networked Microgrids (MG) in electric distribution systems. In practice, the cooperative agent can have limited or no knowledge of the MG asset behavior and detailed models behind the Point of Common Coupling (PCC). This makes the distribution systems unobservable and impedes conventional optimization solutions for the constrained MG power management problem. To tackle this challenge, we have proposed a bi-level RL framework in a price-based environment. At the higher level, a cooperative agent performs function approximation to predict the behavior of entities under incomplete information of MG parametric models; while at the lower level, each MG provides power-flow-constrained optimal response to price signals. The function approximation scheme is then used within an adaptive RL framework to optimize the price signal as the system load and solar generation change over time. Numerical experiments have verified that, compared to previous works in the literature, the proposed privacy-preserving learning model has better adaptability and enhanced computational speed.   
\end{abstract}


\section*{Nomenclature}
\addcontentsline{toc}{section}{Nomenclature}
\begin{IEEEdescription}[\IEEEusemathlabelsep\IEEEsetlabelwidth{$V_1,V_2,V_3$}]
\item[\textbf{Indices}]
\item[$i,j$] Indices of bus numbers $\forall i,j\in \Omega_I$.
\item[$k$] Index of line number $\forall k\in \Omega_K$.
\item[$n$] Index of MG.
\item[$t$] Index of episode/time instant.
\item[\textbf{Parameters}]
\item[$a_f/\ b_f/\ c_f$] Coefficients of the DG quadratic cost function.
\item[$E^{Cap}$] Max. capacity of ESS unit.
\item[$e_{PV},e_{D}$] Prediction error standard deviations.
\item[$G/B$] Real/imag. parts of the bus admittance matrix.
\item[$\hat{I}_{PV}$] Vectors of solar irradiance estimation.
\item[$I^{PV}$] Real normalized solar irradiance.
\item[$P^{Ch/Dis,M}$] Max. ESS charging/discharging limits.
\item[$P/Q^{D}$] Active/reactive load.
\item[$P/Q^{DG,M}$] Max. DG active/reactive power capacity.
\item[$P^{DG,R}$] Max. DG ramp limit.
\item[$P^{PV}$] PV active power output.
\item[$P/Q^{PCC,M}$] Max. active/reactive power flow at the PCCs.
\item[$\hat{P}_{D}$] Vectors of aggregate active load estimation.
\item[$P^{D}$] Real active load.
\item[$Q^{PV,M}$] Max. PV reactive power output limit.
\item[$S$] States in Markov decision process.
\item[$L^{M}$] Max. line flow limit.
\item[$SOC^{M/m}$] Max./min. SOC limits.
\item[$T$] Length of the moving decision window.
\item[$\Delta t$] Time step.
\item[$\alpha/\beta$] Shape parameters of beta distribution.
\item[$\eta_{Ch/Dis}$] Charging/discharging efficiency of ESS unit.
\item[$\lambda^{F}$] Diesel generator fuel price.
\item[$\lambda^{R,M/m}$] Max./min. retail price limits.
\item[$\lambda^{W}$] Wholesale energy price.
\item[$\theta$] Vector of regression parameter.
\item[$\theta^{*}$] Vector of converged regression parameter.
\item[$\theta_{Th}/V_{Th}$] Threshold value.
\item[$\gamma$] Discount factor that defines the preference.
\item[$\delta$] Step size that defines the rate of learning.
\item[$\mu$] Regularization factor.
\item[$\phi$] Forgetting factor.
\item[$\epsilon$] $\epsilon$-greedy exploration factor.
\item[\textbf{Variables}]
\item[$a$] Actions in Markov decision process.
\item[$F$] Fuel consumption of DG.
\item[$SOC$] SOC of the battery system.
\item[$P^{Ch/Dis}$] Charging/discharging power of ESS unit.
\item[$P/Q^{DG}$] DG active/reactive power outputs
\item[$P/Q^{ij}$] Line active/reactive power flows
\item[$P/Q^{PCC}$] Active/reactive power flow at the PCC.
\item[$P^{W}$] Exchanged power with the wholesale market.
\item[$Q^{ESS}$] Reactive power outputs of ESS unit.
\item[$Q^{PV}$] PV inveter reactive power outputs.
\item[$V/\Delta\theta$] Voltage magnitude and phase angle difference.
\item[$x_p/x_q$] MGs power management decision vectors.
\item[$\lambda^R$] Locational retail price signals at the PCCs.
\item[$u^{Ch/Dis}$] ESS charge/discharge binary variables.
\item[\textbf{Functions}]
\item[$Q_t(S,a)$] State-action value function.
\item[$Q^*_t(S,a)$] Optimal state-action value function.
\item[$\hat{Q}_t(S,a|\theta)$] Parameterized approximate state-action value function.
\item[$Q_{S\cdot a}(t|\theta)$] Parameterized regression sub-component with state-action interaction.
\item[$Q_{S}(t|\theta)$] Parameterized regression sub-component with state values.
\item[$Q_{a}(t|\theta)$] Parameterized regression sub-component with action values.
\item[$R(t)$] Reward function in Markov decision process.
\end{IEEEdescription}

%
\IEEEpeerreviewmaketitle

\section{Introduction}
A smart distribution system consisting of networked microgrids (MGs), with local Distributed Generators (DG), Renewable Energy Resources (RES), and Energy Storage Systems (ESS), can facilitate reliable service provision to customers in power systems \cite{SA2013}. Smart independent MGs are considered as a viable solution for electrification of rural areas, which are excluded from traditional electrification programs due to their remote location and financial constraints \cite{Coop_1}. To ensure the long-term sustainability and encourage economic development in rural communities, the feasibility of cooperative business models for rural system electrification has been analyzed previously \cite{Coop_1, Coop_2, Coop_3}. It has been shown that a non-profit cooperative can act as an intermediary agent between the rural MGs and the wholesale market. The power is exchanged between the MGs and the cooperative at a retail rate, and the revenue from electricity sales in the wholesale market is returned to MGs. The retail energy pricing program can be used to influence the MGs' behavior based on the availability of resources. Real cases of cooperative business models with rural MGs as participating members can be found in \cite{Coop_2, Coop_3}. The autonomous cooperative business settings in these cases have been designed to benefit rural communities. 

Coordinating the real-time behavior of multiple privately-owned rural MGs in a cooperative business model is a necessary, yet challenging task \cite{MG_1, Manbachi}. Due to data privacy and ownership concerns for MGs, the main difficulty in the way of obtaining a desirable coordination scheme is the limited access to real-time asset behaviors and models behind the Point of Common Coupling (PCC) with MGs, which hinders conventional model-based constrained power management solvers. This problem becomes more severe as the penetration of MGs in rural distribution systems grows. A wide range of methods have been applied in the literature with the aim of economic operation of the networked MGs, including methods such as heuristic techniques \cite{Heuristic_2018_NM, MMG_NN}, centralized decision models \cite{Centralized_2017_SAA, SOS_2015_DG}, constrained hierarchical control architectures \cite{Centralized_2016_YZhang, Decentralized_ZY, MMG_ZY}, and distributed optimization methods \cite{Decentralized_2017_PKou, Decentralized_2015_DG}.

However, the functionality of previous models \cite{Heuristic_2018_NM, MMG_NN, Centralized_2017_SAA, SOS_2015_DG, Decentralized_2017_PKou, Decentralized_2015_DG, Centralized_2016_YZhang, Decentralized_ZY, MMG_ZY} highly depends on the full system operator's knowledge of MG operation behind the PCC and customers' private data at node-level, including nodal demand load consumption, nodal generation capacities, nodal PV generations, sensitive cost information, asset constraints, as well as MG network topology and configuration data. Access to these information could compromise the data confidentiality and privacy of MGs and customers that participate in a cooperative business setting. Also, previous methods can be mostly categorized as ``model-based'', since the decision agents depend on detailed physical models of the distribution systems. One shortcoming of model-based solutions is their inability to adapt to constantly-changing system conditions when the amount of measurement data is limited. 

A promising alternative to model-based optimization approaches is reinforcement learning (RL), which is a model-free data-driven technique that can be used to optimize the behavior of an agent through repeated interactions with its environment, without full system identification and no \textit{a priori} knowledge of the system. A number of papers have given examples of how RL techniques can be applied in power systems. In \cite{RL_2018_WLiu, RL_2018_EMocanu}, energy consumption scheduling problems were solved for single MGs and individual residential buildings using RL algorithms. However, the above studies only focus on providing optimal solutions to power management problems for single entities instead of addressing coupled decision models for multiple interconnected entities in a cooperative setting.

In this paper, to solve the problem of decision making under incomplete information while providing decision adaptability, a bi-level cooperative framework is proposed using an RL-based method for a distribution system consisting of multiple networked privately-owned MGs: at Level I of the hierarchy, a non-profit cooperative agent maximizes the total MGs' revenue from power exchange with the wholesale market. This is done by setting the locational energy prices, with access only to active/reactive power measurements at the MG PCCs and \textit{aggregate} load and solar irradiance information behind the PCCs. The cooperative agent acts as an intermediary between the MGs and the wholesale market, and returns the revenue to the MGs. At Level II of the hierarchy, each MG Control Center (MGCC) agent receives the price signal from the cooperative agent and solves the power-flow-constrained MG power management problem. The objective at this level consists of the MG operational cost and the allocated revenue from the cooperative agent. In summary, the main contributions of this paper can be listed as follows:
\begin{itemize}
\item The proposed power management system can handle the current limitations raised from data privacy and ownership in the cooperative setting. Considering the model-free nature of our RL-based method, the data privacy of MGs and the data confidentiality of customers are maintained. The power management problem is solved with access to only minimal and aggregated data.
\item The proposed RL solver is faster than conventional optimization solvers since the learned state-action value function acts similar to a \textit{memory} that recalls from the cooperative agent's past experiences to estimate new optimal solutions. This is done by updating the state values at each decision window and without re-solving the decision problem. 
\item The RL framework is trained using a regularized recursive least square methodology with a \textit{forgetting factor}, which enables the decision model to be adaptable against changes in system parameters which are excluded from the cooperative agent's state set.
\end{itemize}

The reminder of the paper is organized as follows: Section \ref{sec:overall} presents the overall decision hierarchy. Section \ref{sec:RL} elaborates the proposed RL-based framework. Section \ref{sec:MG} describes the MG power management problem. Simulation results and conclusions are given in Section \ref{sec:Results} and Section \ref{sec:Con}, respectively.  

\section{Overall Decision Hierarchy}\label{sec:overall}
Fig. \ref{fig.2.1} gives a general overview of the proposed bi-level power management scheme for a distribution system with multiple MGs, as follows:
\begin{figure}
	\vspace{-0pt} 
	\vspace{-0pt}
	\centering
	\includegraphics[width=1.0\linewidth]{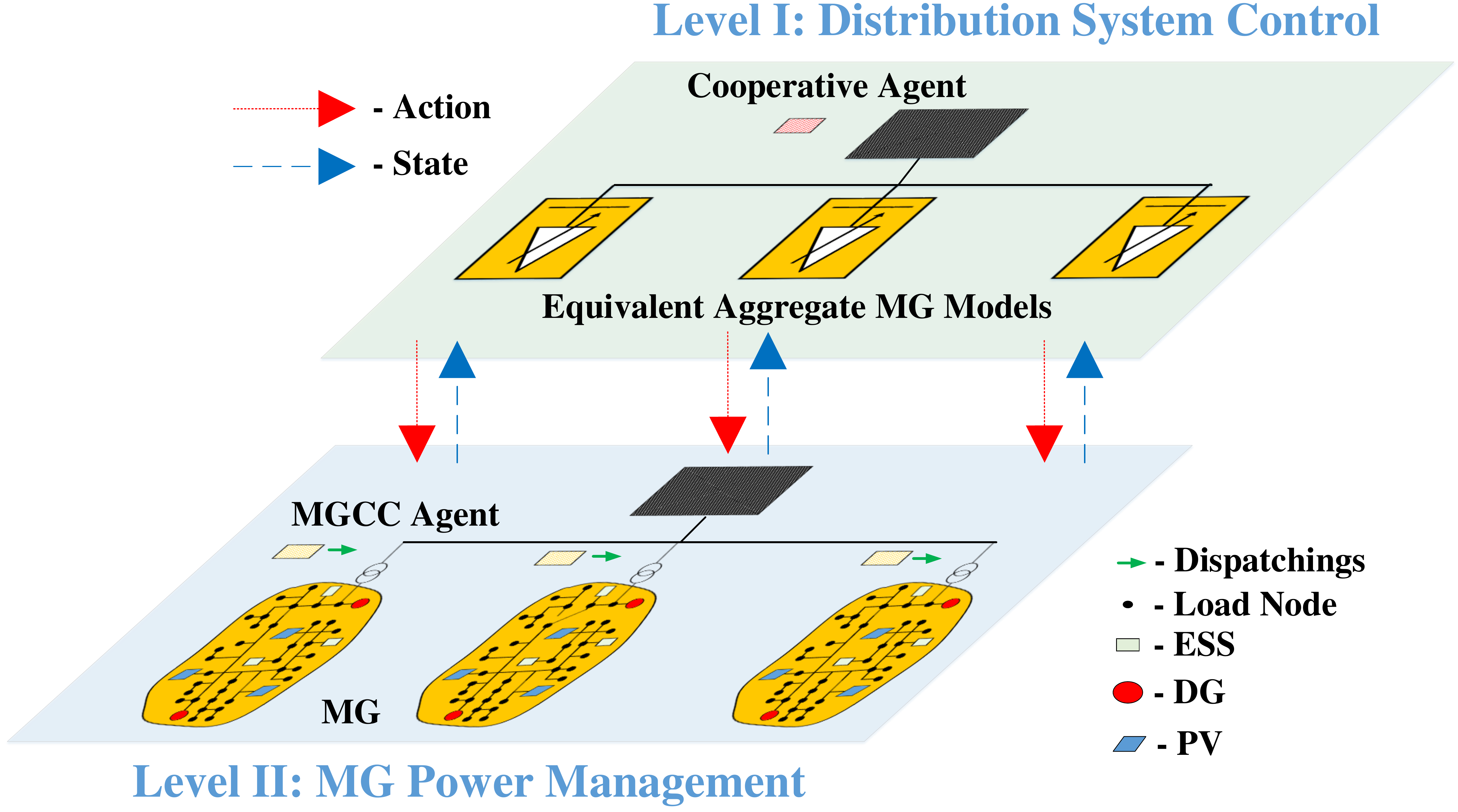}
	\vspace{-20pt} 
	\caption{The architecture of the bi-level networked MGs power management}
	\centering
	\label{fig.2.1}
    \vspace{-15pt} 
\end{figure}

\textbf{Level I - RL-based Distribution System Control:} The non-profit cooperative agent employs an adaptive model-free RL method, developed using a regularized recursive least square function approximation methodology, to find the optimal retail locational price signals for the MGs based on the latest system states. The price signals are then transmitted to MGCC agents. The RL training process is performed by the cooperative agent through repeated interactions with the MGCC agents. At this level, each MG is modeled as an aggregate controllable load which is price-sensitive. The task of the RL algorithm is to discover the complex relationship between locational price and exchanged power with MGs at PCCs, without direct detailed knowledge of system operation behind the PCCs and only with access to estimations of the solar irradiance and aggregate fixed loads for each MG. Based on the definitions of data privacy and confidentiality in smart grid \cite{Data_privacy}, this approach limits the need for access to local cost and operational constraint data of individual MGs in the first place. Hence, the proposed method maintains both the privacy of personal information and privacy of behavior for MGs. Moreover, unlike conventional centralized optimization methods, the proposed RL technique does not need customer confidential information at the node-level, such as customer load consumption, as it only uses aggregate data at the MG PCCs for optimal decision making. Furthermore, renewable and load power uncertainty are represented within the learning model state set. To facilitate adaptive conformation to changes in system parameters that are not included in cooperative agent's state set, such as fuel price, a forgetting mechanism has been integrated into the training process to assign higher importance levels to the latest observed data, compared to previous observations.

\textbf{Level II - MG Power Management:} At the second level, the MGCC agents receive the price signal for a look-ahead moving decision window. Based on the received price signals, each MGCC agent solves a constrained Mixed Integer Nonlinear Programming (MINP) to dispatch their local generation/storage assets to maximize their revenue (or equivalently minimize their cost) in the price-based environment, subject to full AC power flow constraints. Each MG's total revenue includes the cost of operation and the allocated revenue received from the cooperative agent. Based on the solution to this problem, each individual MGCC agent determines the exchanged active and reactive power with the distribution system at PCC.

Note that the RL-based reward maximization problem at Level I is subject to the power-flow-constrained response of MGs at Level II. Since the MGs are sensitive to electricity price, the reward value cannot be maximized by setting the price to its highest value. This will lead to the maximum DG generation, which will result in a decline in the cooperative agent's revenue. Hence, optimal price is reached based on a tradeoff between MGs' over-generation (when price is too high) and over-consumption (when price is too low). Also, note that the response of MGs itself is explicitly constrained by network power flow constraints.

\section{Level I: Adaptive RL-Based Distribution System Control}\label{sec:RL}
At the first level of the hierarchy, a non-profit cooperative agent is in charge of setting the locational price of electricity at different times to maximize the revenue from power exchange with wholesale market, which will be allocated between MGs. This problem is formulated and solved over a moving decision window of length $T$. The difficulty in solving this problem is that the cooperative agent has incomplete knowledge of MGs' asset control and management data. To solve this problem, an RL approach is adopted, in which the decision making cooperative agent observes the response of its environment, consisting of networked MGs, to its actions at different states. Based on the received reward/cost signals from its environment and without explicit modeling, the cooperative agent searches for actions that optimize its expected accumulated received rewards at different system states. 

\subsection{Proposed RL-based Method Strcuture}
A RL framework consists of a Markov decision process including a set of states ($\pmb{S}\in \mathcal{S}$), a set of actions ($\pmb{a}\in \mathcal{A}$), a reward function ($\pi:\mathcal{S}\times\mathcal{A}\rightarrow\mathbb{R}$), and a state-action value function corresponding to each state-action pair ($Q:\mathcal{S}\times\mathcal{A}\rightarrow\mathbb{R}$). These components are defined for the problem at hand, as follows:   

\textit{1) State Set Definition:} In this paper, the system state, which is denoted by $\pmb{S}(t) = (\pmb{S_1}(t),...,\pmb{S_N}(t))^\top$ at time $t$, is a concatenation of MGs' local state vectors ($\pmb{S_n}(t)$ for the $n^{th}$ MG) defined as:
\begin{equation}
\pmb{S_n}(t) = \{\hat{I}_{PV}(t,n),\hat{P}_{D}(t,n)\}
\end{equation}
where, $\hat{I}_{PV}(t,n)$, $\hat{P}_{D}(t,n)$ are the vectors of solar irradiance estimation, and aggregate active load power estimation for the $n^{th}$ MG at time $t$, respectively. Hence, to define the global state, the cooperative agent needs to estimate or predict the uncertain aggregate solar irradiance and load at the PCC for each MG. To represent the uncertainty of the prediction process, prediction error values are considered to the actual underlying solar irradiance and load values, as shown below:
\begin{subequations}
\begin{equation}
\hat{I}_{PV}(t,n) \sim Beta(\alpha,\beta)
\end{equation}
\begin{equation}
\alpha = \frac{\beta(\sum_i I_{i,t,n}^{PV})}{(1-\sum_i I_{i,t,n}^{PV})}
\end{equation}
\begin{equation}
\beta = (1-\sum_i I_{i,t,n}^{PV})(\frac{\sum_i I_{i,t,n}^{PV}(1+\sum_i I_{i,t,n}^{PV})}{e^{2}_{PV}}-1)
\end{equation}
\begin{equation}
\hat{P}_{D}(t,n) \sim \mathcal{N}(\sum_i P_{i,t,n}^{D},e^{2}_{D}(t)) 
\end{equation}
\end{subequations}
where, $\sum_i I_{i,t,n}^{PV}$  and $\sum_i P_{i,t,n}^{D}$ are the real aggregate normalized solar irradiance and load over the decision window, and $e_{PV}$ and $e_{D}$ are the beta and Gaussian estimation error standard deviations. The values of parameters of beta and Gaussian distributions are adopted from the \cite{Solar_beta,Solar_beta2,Load_Wu}. 

\textit{2) Action Set Definition:} Given the definition of model states, the global action vector is similarly defined by the locational retail price signals at the PCCs with MGs, denoted as $\lambda_{t,n}^R$ for the $n^{th}$ MG, $\pmb{a}(t) = (\lambda_{t,1}^R,...,\lambda_{t,N}^R)^\top$. 

\textit{3) Reward Function Definition:} The reward function at time $t$ represents the discounted accumulated revenue of the cooperative agent over the moving decision window with length $T$:
\begin{equation}
\label{eq:Areward}
R(t) = \sum_{t'=0}^{T-1}\gamma^{t'}(\lambda^{W}_{t+t'}P^{W}_{t+t'} - \sum_{n=1}^{N}\lambda^{R}_{t+t',n}P^{PCC}_{t+t',n})
\end{equation}
where, $\gamma$ is a discount factor ($0\leq \gamma \leq 1$) that defines the cooperative agent's preference for the immediate reward, defined as the revenue at time $t$, $\pi(t) = \lambda^{W}_{t}P^{W}_{t} - \sum_{n=1}^{N}\lambda^{R}_{t,n}P^{PCC}_{t,n}$. Also, $\lambda^{W}_{t}$ denotes the wholesale energy price, $P^{W}_t$ is the exchanged power with the wholesale market, where $P^{W}_t \leq 0$ represents power import from the wholesale market. $P^{PCC}_{t,n}$ is the active power transfer between grid and the $n^{th}$ MG through the PCC, where $P^{PCC}_{t,n}\geq 0$ implies export from MGs to grid. The extreme case of $\gamma=0$ represents a myopic cooperative agent, which favors only the immediate economic rewards and assigns zero weights to future expected rewards. However, as the discount factor increases the cooperative agent starts to include future expected rewards into its optimal decision problem. Hence, when the discount factor reaches $\gamma=1$ the cooperative agent assigns equal weights to all the expected reward values for all the time instants in the decision window. 

\textit{4) State-action Value Function Parameterization:} To optimize the cooperative agent's action, an auxiliary state-action value function is formed, denoted as $Q(S,a)$, which can be thought of as a replacement for the explicit system model. The state-action value function determines the long-term accumulated expected reward given the current state and action vectors:
\begin{equation}
Q_t(\pmb{S},\pmb{a}) = E\{\sum_{t' = 0}^{ T-1}\gamma^{t'}\pi(t + t')|\pmb{S}(t)=\pmb{S},\pmb{a}(t)=\pmb{a}\}\label{eq_Q}
\end{equation}
where, $Q_t(\pmb{S},\pmb{a})$ is the expected accumulated reward if the initial starting state is $\pmb{S}(t)$, while the selected initial action is $\pmb{a}(t)$, and the latest optimal policy is followed for every other time-step in the future. The expectation operator $E\{\}$ is calculated with respect to the future expected action-states, which in this case are in turn functions of the solar-load uncertain powers. 

The goal of RL is to learn an optimal state-action value function, $Q^*_t(\pmb{S},\pmb{a})$, that satisfies the Bellman optimality equation \cite{RL_2017_RSS}, as follows:
\begin{equation}
\label{eq:bellman}
Q^*_t(\pmb{S},\pmb{a}) = E\{\pi(t+1) + \gamma\cdot \max_{\pmb{a'}}Q^*_t(\pmb{S}(t+1),\pmb{a'})\}
\end{equation}

Since solving (\ref{eq:bellman}) directly is not possible, RL provides a framework to obtain the optimal state-action value function which satisfies (\ref{eq:bellman}) using an iterative episodic learning environment. To implement this framework for the cooperative agent interacting with multiple MGs, the state-action value function is parameterized employing a multivariate polynomial regression approximation technique \cite{RL_2017_RSS,poly_2,poly_3}, defined by $\hat{Q}_t$, which consists of three multivariate polynomial elements with maximum degree 2:
\begin{equation}
\label{eq:approx}
Q_t(\pmb{S},\pmb{a}) \approx \hat{Q}_t(\pmb{S},\pmb{a}|\pmb{\theta}) = Q_{\pmb{S}\cdot\pmb{a}}(t|\pmb{\theta}) + Q_{\pmb{S}}(t|\pmb{\theta}) + Q_{\pmb{a}}(t|\pmb{\theta})
\end{equation}

Given the regression parameter vector $\pmb{\theta}$, $Q_{\pmb{S}\cdot\pmb{a}}$, $Q_{\pmb{S}}$, and $Q_{\pmb{a}}$ are the parameterized sub-components that quantify the impacts of state-action interaction $Q_{\pmb{S}\cdot\pmb{a}}(t|\pmb{\theta})$, state values $Q_{\pmb{S}}(t|\pmb{\theta})$, and action values $Q_{\pmb{a}}(t|\pmb{\theta})$, respectively. These regression sub-components in multivariate polynomial regression model are defined as follows: 
\begin{equation}
\label{eq:sub1}
Q_{\pmb{S}\cdot\pmb{a}}(t|\pmb{\theta})= \sum_{n=1}^{N}\theta^1_{t,n}\lambda^R_{t,n}\hat{I}_{PV}(t,n) + \sum_{n=1}^{N}\theta^2_{t,n}\lambda^R_{t,n}\hat{P}_{D}(t,n)
\end{equation}
\begin{equation}
Q_{\pmb{S}}(t|\pmb{\theta})= \sum_{n=1}^{N}\theta^3_{t,n}\hat{I}_{PV}(t,n) + \sum_{n=1}^{N}\theta^4_{t,n}\hat{P}_{D}(t,n)\label{eq:sub2}
\end{equation}
\begin{equation}
Q_{\pmb{a}}(t|\pmb{\theta})= \sum_{n=1}^{N}\theta^5_{t,n}\lambda^R_{t,n} + \theta^6\label{eq:sub3}
\end{equation}
where, $\pmb{\theta}=\{\theta^k_{t,n},\theta^k\}$ constitute the parameters of the approximate state-action value function that have to be learned by the cooperative agent through repeated interaction with the MGs. 

Together these three components form a bilinear regression model to parametrize the state-action value function (i.e., the regression model is linear with respect to each of its arguments.) The reason for selecting a bilinear regression model is the structure of the reward function \eqref{eq:Areward}, which also follows a bilinear relationship between the price signal and the aggregate power measured at MG PCCs and the substation. Furthermore, the state-action value parameterization shown in \eqref{eq:sub1}-\eqref{eq:sub3} offers two critical advantages compared to other types of function approximators: 1) Using a bilinear regression model will simplify optimal action selection procedure considerably, as will be shown in Section \ref{action}. For instance, if an artificial neural network is used, optimal action selection becomes intractable. However, using the proposed bilinear regression model, optimal action selection reduces to linear programming, which can be solved easily. 2) A basic challenge in choosing the form of a function approximator is the tradeoff between over-parametrization and estimation accuracy. For example, as we increase the degree of the multivariate polynomial approximator the value estimation accuracy for new state-action pairs would also improve; however, at some point the function approximator becomes over-parameterized and will start overfitting to the available data, at which point the performance declines. We observed that by limiting the degree of the multivariate polynomial degree to 2, the best estimation accuracy can be achieved while maintaining a safe margin to avoid overfitting under various practical case studies.

\subsection{Adaptive RL-based Method Training}\label{action}
To achieve this task we have adopted an adaptive episodic learning mechanism, which is shown in Fig. \ref{fig.4.1}. Each episode in the learning process corresponds to an online decision instant. Hence, as the decision window rolls along time new episodes are perceived by the cooperative agent. The learning process has the following steps:
\begin{figure}
	\vspace{-0pt} 
	\vspace{-0pt}
	\centering
	\includegraphics[width=1.0\linewidth]{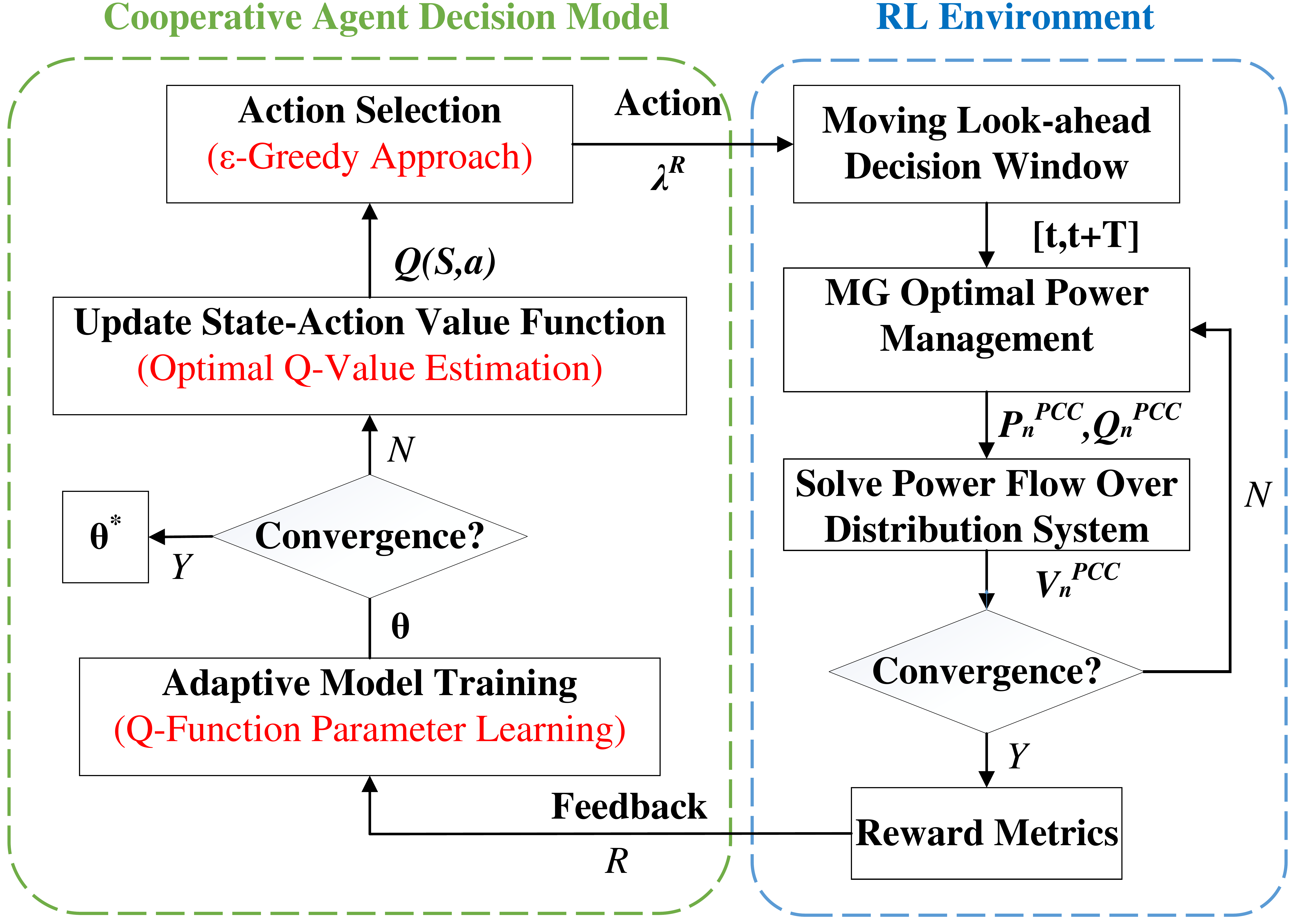}
			\vspace{-20pt}\
	\caption{Proposed RL-based framework}  
	\centering
	\label{fig.4.1}
				\vspace{-20pt}\
\end{figure}

\textbf{Step 1.} \textit{Initialization:} The time index is initialized as $t = t_0$, representing the first episode. The parameters of the state-action value function are initialized, $\pmb{\theta} \leftarrow \pmb{\theta}(t_0)$. The initial state of the system, corresponding to solar irradiance and aggregate load of all the MGs for the decision window $[t_0,t_0 + T]$ is predicted, $\pmb{S}(t_0), ..., \pmb{S}(t_0+T)$. Note that these predicted states, while representing system uncertainty, are updated continuously as the decision window rolls along time. 

\textbf{Step 2.} \textit{$\epsilon$-greedy Action Selection:} Based on the latest state-action value function defined by parameter $\pmb{\theta}$, the optimal actions are estimated for the decision window $[t,t+T]$ to maximize the cooperative agent's accumulated reward, as follows:
\begin{equation}
\label{eq:optaction}
\begin{split}
\pmb{a_{opt}}&(t') = \argmaxA_{\pmb{a'}}Q_{t'}(\pmb{S}(t'),\pmb{a'})\\
s.t.&\ \ \pmb{a'} = (\lambda_{t',1}^R,...,\lambda_{t',N}^R)^\top\\
\lambda^{R,m} \leq & \lambda_{t',i}^R \leq \lambda^{R,M}, \forall\ i = \{1,...,N\}\\
&\forall t' = \{t,...,t+T\}
\end{split}
\end{equation}
where, $\pmb{\rho_\lambda} = [\lambda^{R,m},\lambda^{R,M}]$ defines the minimum/maximum range of action for retail price. Note that given the parameterization for $Q_t(\pmb{S},\pmb{a})$ in (\ref{eq:sub1})-(\ref{eq:sub3}), (\ref{eq:optaction}) is basically a set of linear programs, which can be solved efficiently using off the shelf solvers. A critical aspect of (\ref{eq:optaction}) is that the obtained optimal action, $\pmb{a_{opt}}(t)$, is calculated with respect to the latest state-action value function, which could be far from being accurate in the early stages of training. Hence, to reduce the risk of sub-optimality and to strike a balance between exploration and exploitation of decision space, an $\epsilon$-greedy action selection method \cite{RL_2017_RSS} is adopted, with $0\leq\epsilon\ll1$, to select the cooperative agent's action at time $t$: 
\begin{equation}
\label{eq:actselect}
\begin{split}
\pmb{a}(t) &= \begin{cases}
    \pmb{a_{opt}}(t)& \text{if } r\geq\epsilon\\
    \lambda_{t,i}^R\sim U\{\pmb{\rho_\lambda}\}\ \forall i & \text{if } r<\epsilon\\
\end{cases}
\end{split}
\end{equation}
where, $r$ is a random number selected uniformly, $r\sim U\{[0,1]\}$, with U\{\pmb{A}\} representing uniform probability distribution over the set $\pmb{A}$. The randomization \eqref{eq:actselect} promotes continuous exploration of action space to improve the outcome of the learning process. Upon obtaining the action vector $\pmb{a}(t)$, retail price signals are sent to each MGCC agent. 

\textbf{Step 3.} \textit{Networked MG Power Management:} Based on the received price signals, $\lambda_{t',n}^R, \forall n, t' = \{t,...,t+T\}$, each MGCC agent solves its optimal power management problem (Section IV). Based on the solutions at this stage, the aggregate power injection/withdrawal to/from the grid are obtained at the PCCs with the MGs, denoted as $P_{t',n}^{PCC}$ and $Q_{t',n}^{PCC}$, $\forall n, t' = \{t,...,t+T\}$.

\textbf{Step 4.} \textit{Accumulated Reward Calculation:} Based on the outcomes of the MG power managements, the net power exchange with the wholesale market, $P_t^W$, is determined and used to calculate the discounted accumulated revenue for the decision window $[t,t+T]$, using (\ref{eq:Areward}).

\textbf{Step 5.} \textit{Adaptive Model Training:} Using the observed reward signal, the regression models defined in (\ref{eq:sub1})-(\ref{eq:sub3}) are updated, based on a gradient descent approach to modify the parameters in the direction of improving the generalization capacity of the state-action value function \cite{RL_2017_RSS}:
\begin{equation}
\label{eq:gd}
\pmb{\theta}(t+1) \leftarrow \pmb{\theta}(t) + \delta\{R(t) - \hat{Q}_t(\pmb{S},\pmb{a}|\pmb{\theta})\}\nabla_{\pmb{\theta}}\hat{Q}_t(\pmb{S},\pmb{a}|\pmb{\theta})
\end{equation}
where, $\delta$ is the step size that defines the rate of learning. Note that ideally we require $\hat{Q}_t(\pmb{S},\pmb{a}|\pmb{\theta})=R(t)$, which implies that the approximate state-action value function is able to accurately predict the accumulated reward. Accordingly, (\ref{eq:gd}) is devised to reduce this prediction error over time. To implement (\ref{eq:gd}), two points have to be taken under consideration: 1) since data acquisition and the training process both depend on cooperative agent action selection, approximate RL algorithms are known to be prone to overfitting and over-estimation of the values of state-action pairs \cite{Smart2000}. Hence, a regularization mechanism has to be adopted to reduce the risk of overfitting, 2) the distribution system parameters are subject to change over time. These time-varying parameters, such as price of fuel, are not directly captured in the Markov decision process's state definition. This makes the learned model susceptible to failure in case considerable changes occur in the values of these parameters. Hence, the training process needs to be \textit{adaptive} to enable cooperative agent to quickly conform to new system conditions. To implement (\ref{eq:gd}) while considering the above-mentioned points, a regularized recursive least squares algorithm with exponential forgetting is designed \cite{Houtzager2012}. The regression parameters are updated recursively, as follows:
\begin{equation}
\label{eq:update1}
\pmb{\theta}(t+1) \leftarrow \pmb{\theta}(t) + \pmb{\Delta}(t)\pmb{x}(t)\{R(t) - \hat{Q}_t(\pmb{S},\pmb{a}|\pmb{\theta})\}
\end{equation}
\begin{equation}
\label{eq:update2}
\pmb{\Delta}(t+1) \leftarrow \hat{\pmb{\Delta}}(t+1)(I + \mu\hat{\pmb{\Delta}}(t+1))^{-1}
\end{equation}
\begin{equation}
\label{eq:update3}
\hat{\pmb{\Delta}}(t+1) \leftarrow \frac{1}{1-\phi}(\pmb{\Delta}(t) - \frac{\pmb{\Delta}(t)\pmb{x}(t)\pmb{x}(t)^\top\pmb{\Delta}(t)}{1 + \pmb{x}(t)^\top\pmb{\Delta}(t)\pmb{x}(t)})
\end{equation}
where, $\pmb{x}(t) = (\pmb{S}(t),\pmb{a}(t))^\top$ represents the latest cooperative agent's observation, $\pmb{\Delta}$ is an auxiliary matrix mimicking the regression pseudo-inverse matrix, $\mu$ is the regularization factor which is used for re-scaling the model covariance, and $0\leq\phi < 1$ is the forgetting factor. The regularization factor acts as a weight for penalizing the Euclidean norm of parameter vector (i.e., $||\pmb{\theta}||_2$) in a ridge regression setting to prevent overfitting. The forgetting factor enables the cooperative agent to ``forget'' its earlier experiences in favor of the newer observations by assigning lower weights to the previously learned parameters. Hence, the forgetting factor introduces an exponential extenuation of data history over time. 

\textbf{Step 6.} \textit{State Transition:} The decision window is moved forward to the new episode, $t\leftarrow t+1$. The new system state for the decision window, $[t,t+T]$ is predicted and denoted as $\{\pmb{S}(t), ..., \pmb{S}(t+T)\}$. 

\section{Level II: MGCC Agent Power Management}\label{sec:MG}
At Level II, each MG receives the price signals from the cooperative agent to solve the constrained optimal power management problem within a moving decision window individually, as shown in the paper Appendix, \eqref{eq3_1}-\eqref{eq3_15}. Each MG is comprised of local DGs, ESS, solar Photo-Voltaic (PV) panels and a number of loads. Hence, to account for the impacts of MGs on each other, the MG-level optimal power flow solver is based on an interactive non-linear programming algorithm. The steps of the interactive power flow solution are as follows:

\textbf{Step I.} \textit{Receive input signals from Level I}: The MGs receive the locational retail price signals at the PCCs, $\lambda_{t,n}^R$, from the cooperative agent. 

\textbf{Step II.} \textit{Solve individual MG optimal power management problem}: Given $\lambda_{t,n}^R$ and the estimated voltage at PCC, the power management problem \eqref{eq3_1}-\eqref{eq3_15} is solved independently by each MGCC, and the exchanged active and reactive powers at the PCCs are obtained for each MG. 

\textbf{Step III.} \textit{Solve power flow problem over distribution system}: Treating MGs as fixed PQ loads in the external distribution system, power flow is solved over the network connecting the MGs. The total substation exchanged power, $P^{W}_{t}$, and voltage values at PCCs, $V^{PCC}_{t,n}$, are updated based on the power flow solution.

\textbf{Step IV.} \textit{Check convergence}: Go back to Step III to update PQ values corresponding to each MG, until the changes in voltage values at MG PCCs are smaller than a threshold value $V_{Th}$.

To summarize, the pseudo-code of the proposed bi-level RL-based framework has been shown in Algorithm \ref{alg:Bi_level}.
\begin{algorithm}
\caption{Bi-level RL-based power management method}\label{alg:Bi_level}
\begin{algorithmic}[1]
\State {Select $T, \gamma, \delta, \mu, \phi, \epsilon,\pmb{\theta}(t_0)$}
\Procedure{Level I: RL Action Selection}{$\pmb{\theta}$}
    \State {$t \leftarrow 1$}
    \State {$\pmb{S} \leftarrow [\pmb{S}(t), ..., \pmb{S}(t+T)]$}
    \State {$Q_t(\pmb{S},\pmb{a}) \leftarrow \hat{Q}_t(\pmb{S},\pmb{a}|\pmb{\theta})$}
    \State $\pmb{a_{opt}}(t) \leftarrow$ Solve linear program (10)
    \State $\lambda_{t,i}^R\sim U\{\pmb{\rho_\lambda}\}$
    \State {$r\sim U\{[0,1]\}$} 
    \If {$r\geq\epsilon$}
    \State $\pmb{a}(t) \leftarrow \pmb{a_{opt}}(t)$
    \Else
        \State $\pmb{a}(t) \leftarrow \lambda_{t,i}^R$
    \EndIf
\EndProcedure

\Procedure{Level II: MGCC Agent Power Management}{$\pmb{a}$}
    \State $k \leftarrow 1$ 
    \State {$\lambda^{R} \leftarrow \pmb{a}(t)$, $V_n(k) \leftarrow V^{PCC}_{t,n}$}
    \State {$P^{PCC}_{t,n},Q^{PCC}_{t,n} \leftarrow$ Solve \eqref{eq3_1}-\eqref{eq3_15} $\forall n$ with $V_n(k)$}
    \State {$V_n(k) \leftarrow$ Solve power flow with $\{{P^{PCC}_{t,n},Q^{PCC}_{t,n}}\}$}
   \If {$\Delta |V_n| \geq V_{Th}$}
    \State $k\leftarrow k+1$ 
    \State Go back to Step 18
    \Else
    \State Go to Step 27
   \EndIf
\EndProcedure

\Procedure{Level I: RL Update State-Action Value Function}{$P^{PCC},P^{W},\pmb{S},\pmb{a},\pmb{\theta}$}
    \State {$R(t) \leftarrow \sum_{t'=0}^{T-1}\gamma^{t'}(\lambda^{W}_{t+t'}P^{W}_{t+t'} - \sum_{n=1}^{N}\lambda^{R}_{t+t',n}P^{PCC}_{t+t',n})$}
    \State {$\hat{Q}_t(\pmb{S},\pmb{a}|\pmb{\theta}) \leftarrow Q_{\pmb{S}\cdot\pmb{a}}(t|\pmb{\theta}) + Q_{\pmb{S}}(t|\pmb{\theta}) + Q_{\pmb{a}}(t|\pmb{\theta})$}
    \State {$\pmb{\theta}(t+1) \leftarrow \pmb{\theta}(t) + \delta\{R(t) - \hat{Q}_t(\pmb{S},\pmb{a}|\pmb{\theta})\}\nabla_{\pmb{\theta}}\hat{Q}_t(\pmb{S},\pmb{a}|\pmb{\theta})$}
   \If {$||\pmb{\theta}(t+1)-\pmb{\theta}(t)|| \geq \theta_{Th}$}
     \State {$t\leftarrow t+1$}
     \State Go back to Step 4
    \Else
    \State $\theta^{*} \leftarrow \theta(t+1)$
    \State Output $\theta^{*}$
   \EndIf
\EndProcedure
\end{algorithmic}
\end{algorithm}
	
\section{Numerical Results}\label{sec:Results}
\begin{figure} [h]
	\vspace{-0pt} 
	\vspace{-0pt}
	\centering
	\includegraphics[width=0.85\linewidth]{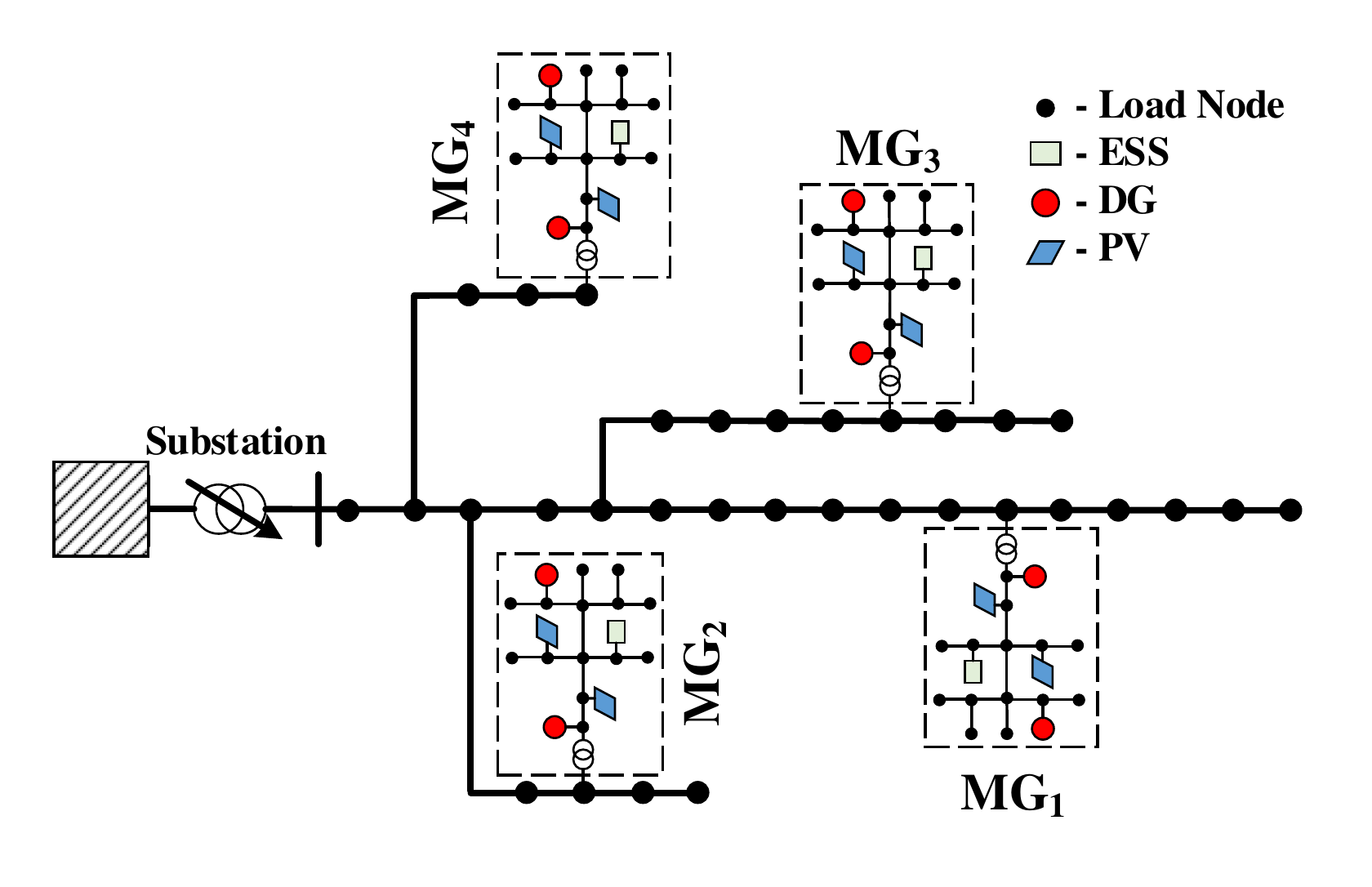}
	\vspace{-15pt}\
	\caption{Test system under study.}  
	\centering
	\label{fig.5.00}
		\vspace{-15pt}\
\end{figure}

The proposed method is tested on a modified medium voltage 33-bus distribution network \cite{33bus_1989_ME}, which has been widely used for studies pertaining to distribution system \cite{RS_33BUS}. The case study consists of four MGs as shown in Fig. \ref{fig.5.00}. Each MG is modeled as a modified IEEE 13-bus network at a low voltage level \cite{13_bus}. Hence, the system has a total number of 85 nodes. To represent a realistic model, we simulated an unbalanced system, where the loads and generators are almost uniformly distributed across phases. Note that the proposed model-free power management technique applies to both balanced and unbalanced systems. Table \ref{table_RL_Parm} presents all setting parameters for the proposed RL-based method in this paper.

\begin{table}[]
		\centering
		\renewcommand{\arraystretch}{1.3}		
		\caption{RL-based Method Parameters}
		\label{table_RL_Parm}
			\vspace{-5pt}\
		\begin{tabular}{ccccccc}
			\hline
			\multicolumn{1}{c}{Parameters}   & \multicolumn{1}{c}{$\gamma$}   & \multicolumn{1}{c}{$\delta$} & \multicolumn{1}{c}{$\mu$}    & \multicolumn{1}{c}{$\phi$} & \multicolumn{1}{c}{$\epsilon$} \\ \hline
			\multicolumn{1}{c}{Values} & \multicolumn{1}{c}{0.99}  & \multicolumn{1}{c}{0.01} & \multicolumn{1}{c}{$1\times10^{-5}$} & \multicolumn{1}{c}{0.01}  & \multicolumn{1}{c}{0.1}\\ \hline
		\end{tabular}
\end{table}

\subsection{System Operation Outcomes}
\begin{figure}
\centering
\subfloat[Aggregate active load profile of the MGs \label{case2load}]{
\includegraphics[width=0.8\linewidth]{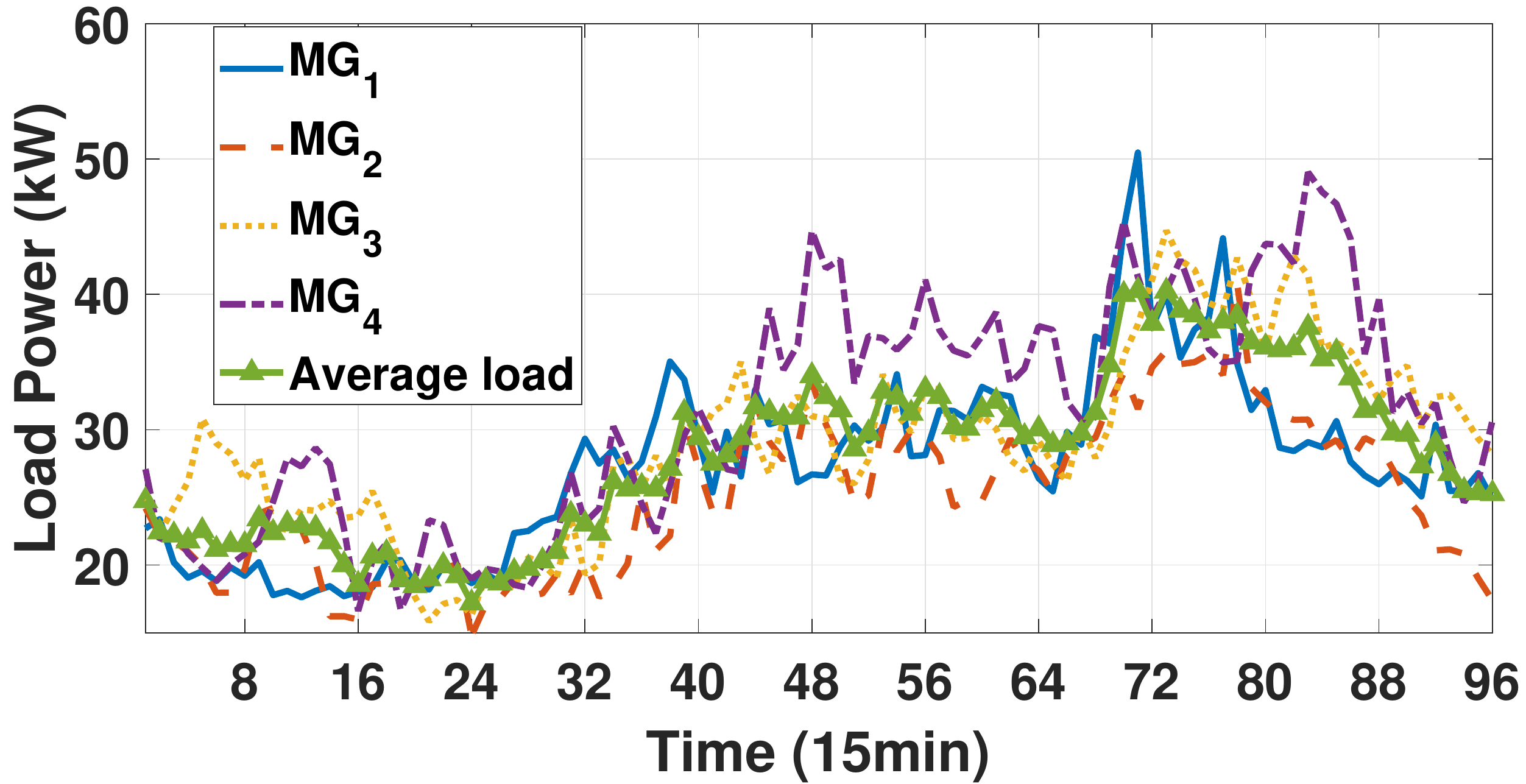}
}
\hfill
\subfloat[Aggregate PV power of the MGs \label{case2PV}]{
\includegraphics[width=0.8\linewidth]{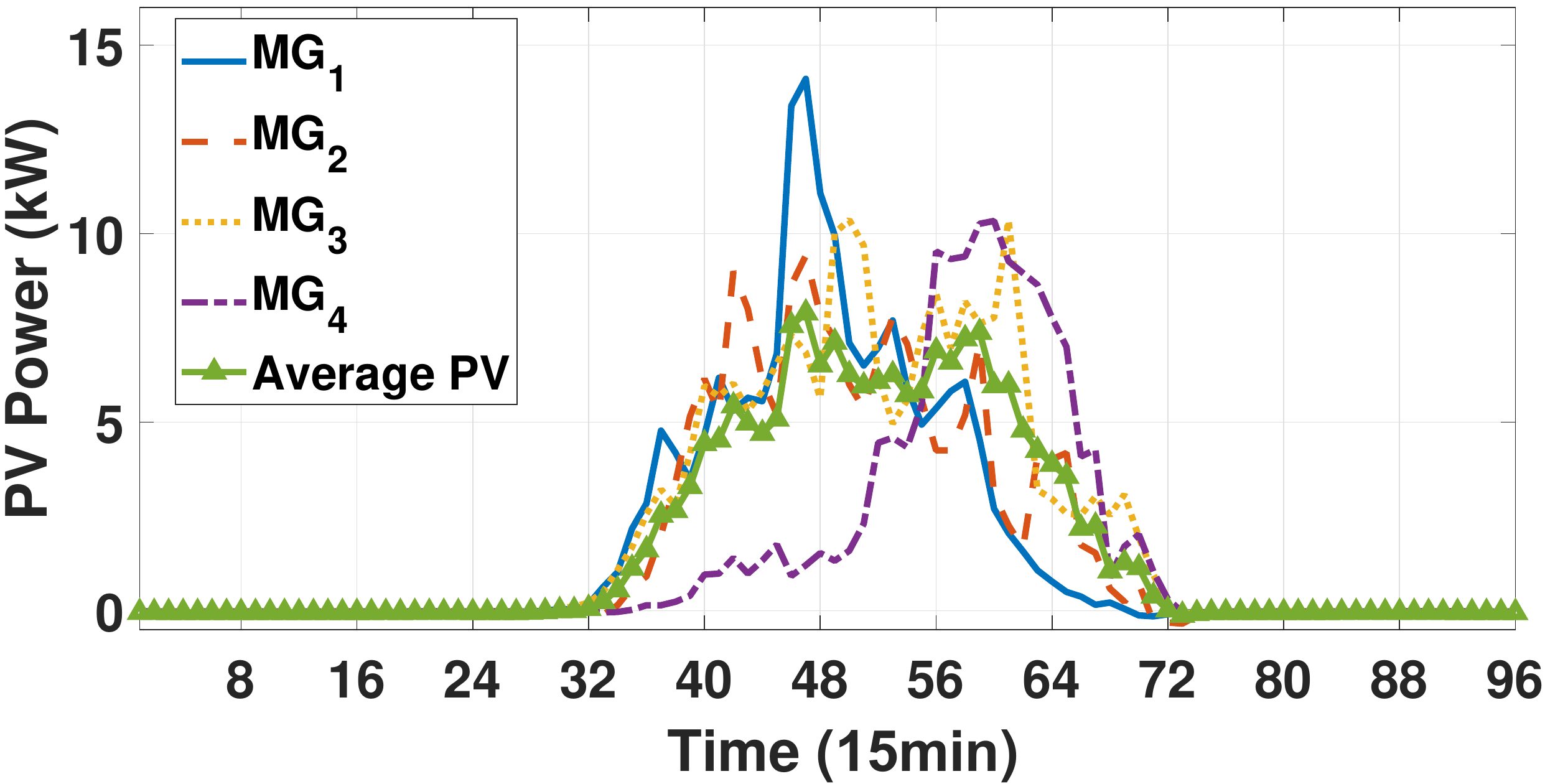}
}
\hfill
\subfloat[Wholesale market price \label{case2Price}]{
\includegraphics[width=0.82\linewidth]{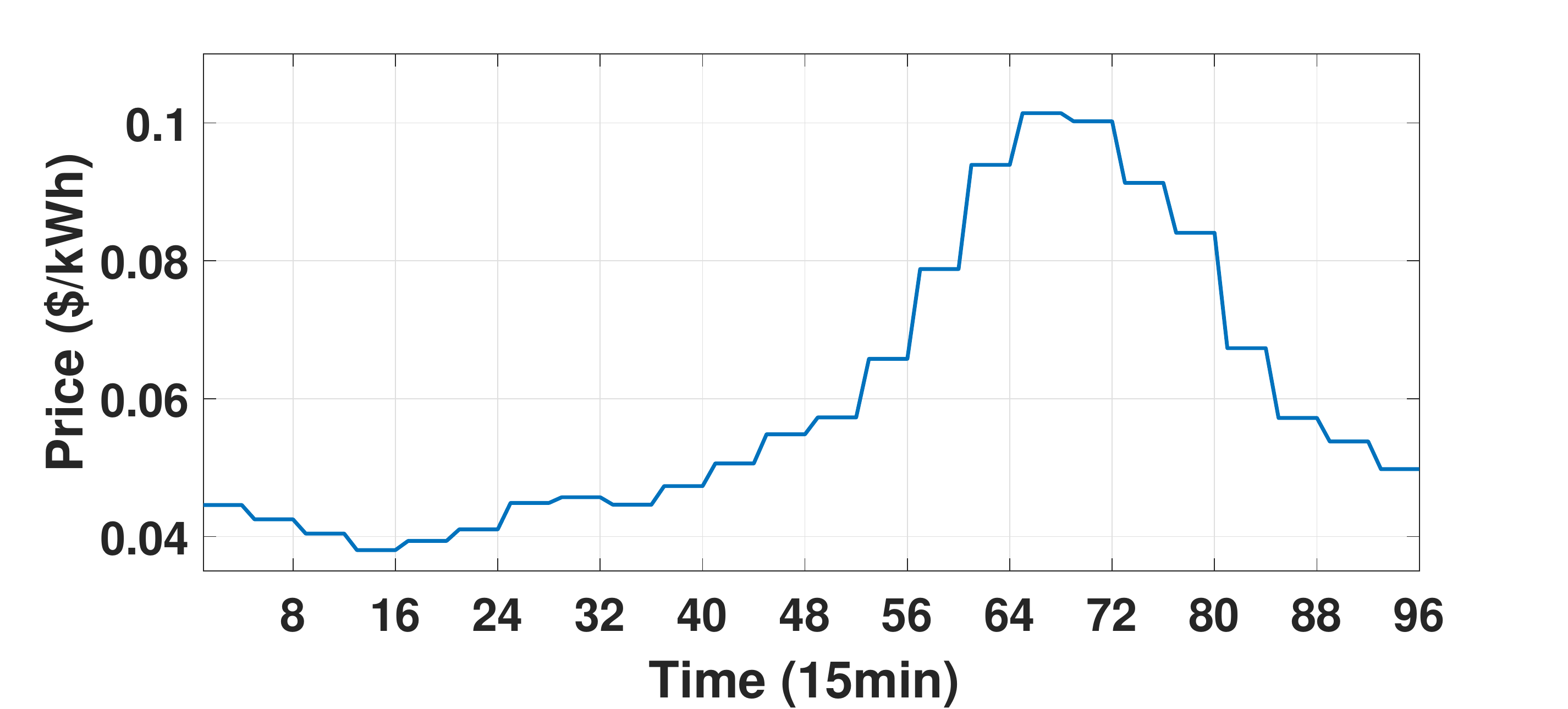}
}
\caption{Input data for the case study}
\label{fig.6.1}
\end{figure}

The \textit{aggregate} active load profiles of all the MGs and the average load are presented in Fig. \ref{case2load}. The \textit{aggregate} solar active generations in each MGs have been shown in Fig. \ref{case2PV}. Both load demands and PV generations data with 15 minutes time resolution are obtained from smart meters to provide realistic numerical experiments. The wholesale market prices used in the numerical case study have been shown in Fig. \ref{case2Price}, which are adopted from the historical wholesale electricity market data from U.S. Energy Information Administration \cite{EIA}.

\begin{figure}
	\vspace{-0pt} 
	\vspace{-0pt}
	\centering
	\includegraphics[width=0.85\linewidth]{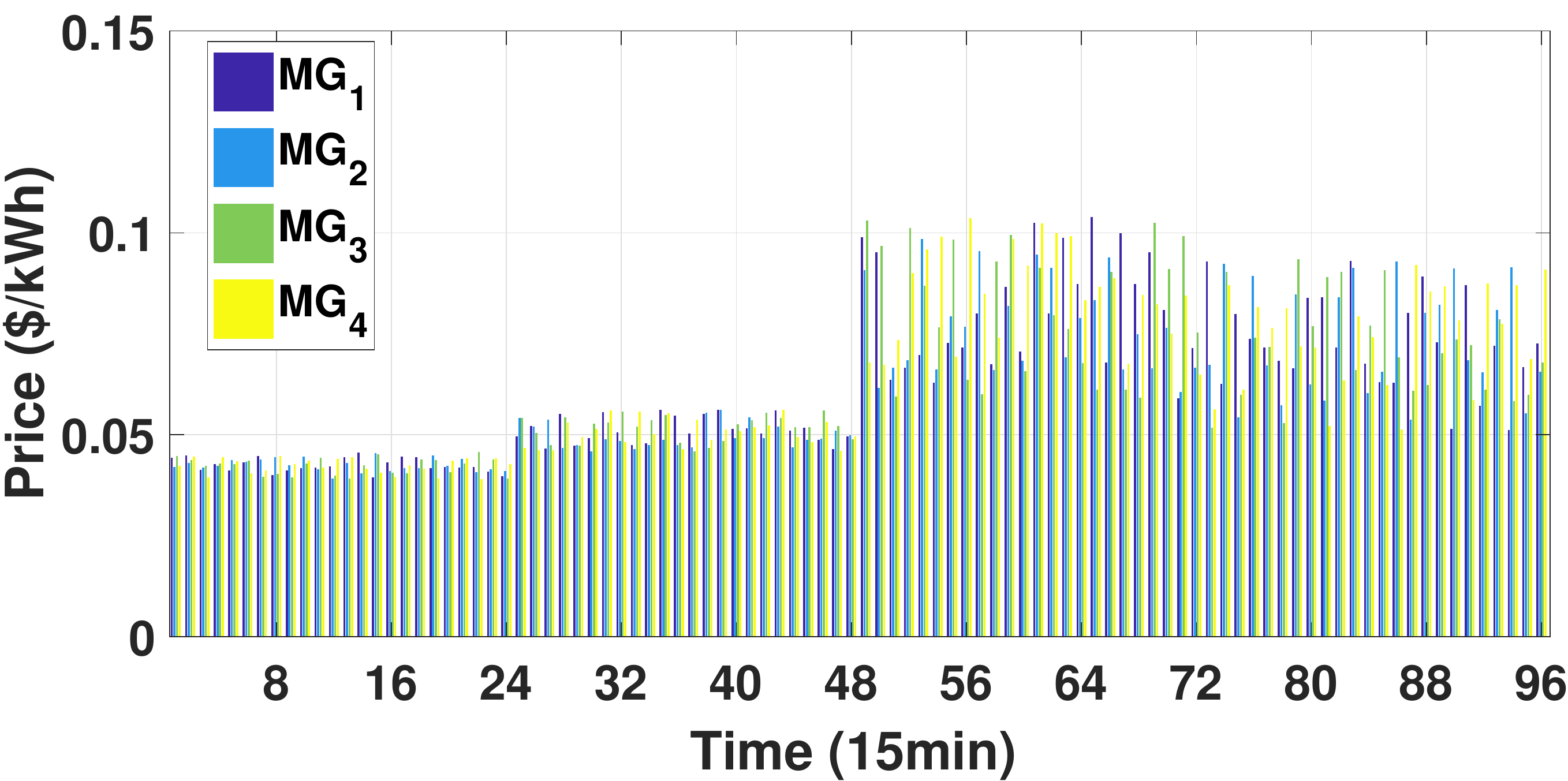}
	\vspace{-5pt}\
	\caption{Optimal locational retail price signals (Level I actions)}  
	\centering
	\label{fig.6.2}
		\vspace{-10pt}\
\end{figure}

\begin{figure}
	\vspace{-0pt} 
	\vspace{-0pt}
	\centering
	\includegraphics[width=0.85\linewidth]{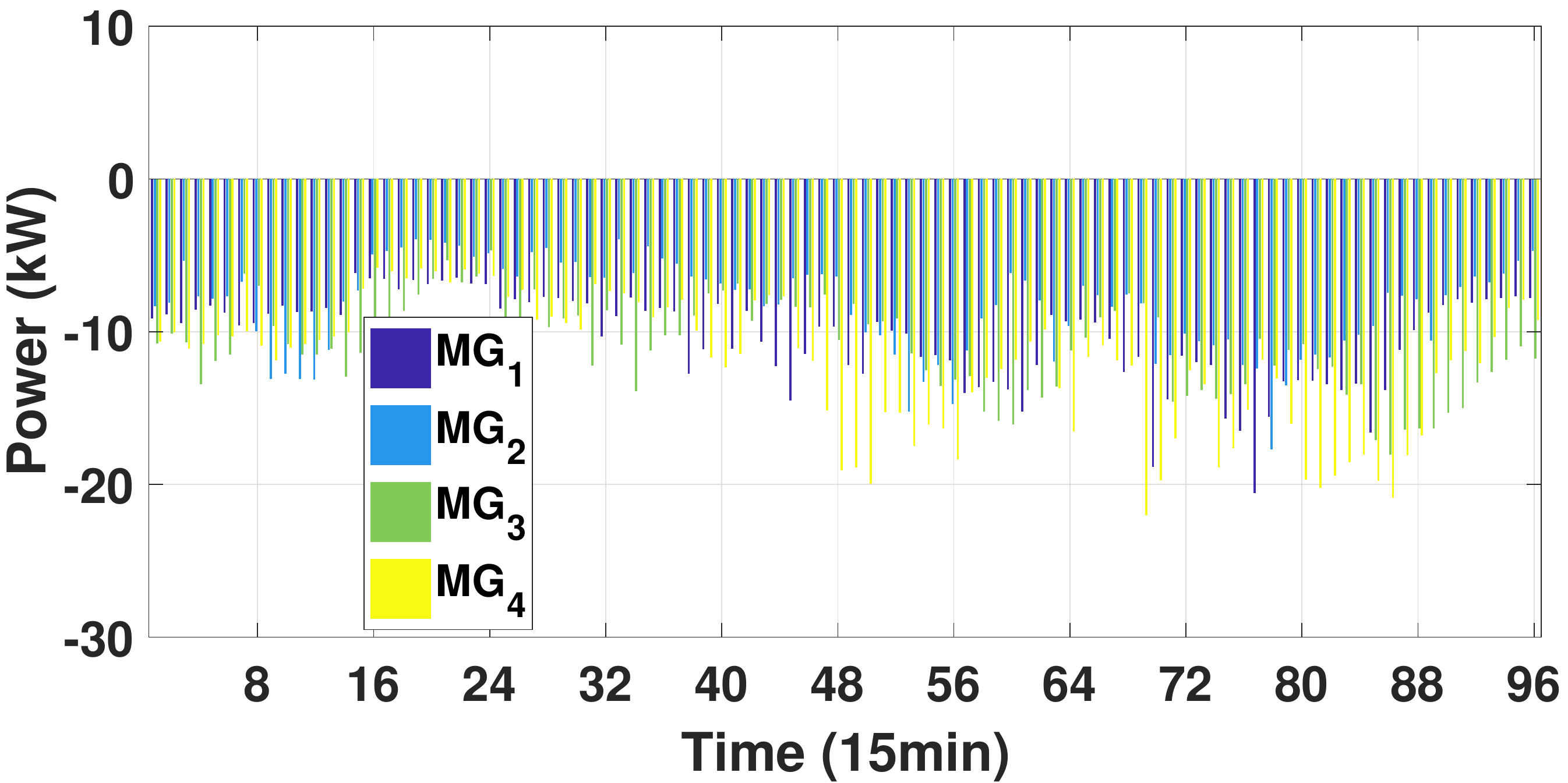}
	\vspace{-5pt}\
	\caption{Optimal power transfer through PCC of MGs (Level II responses to optimal actions)}
	\centering
	\label{fig.6.3}
		\vspace{-10pt}\
\end{figure}

\begin{figure}
	\vspace{-0pt} 
	\vspace{-0pt}
	\centering
	\includegraphics[width=0.85\linewidth]{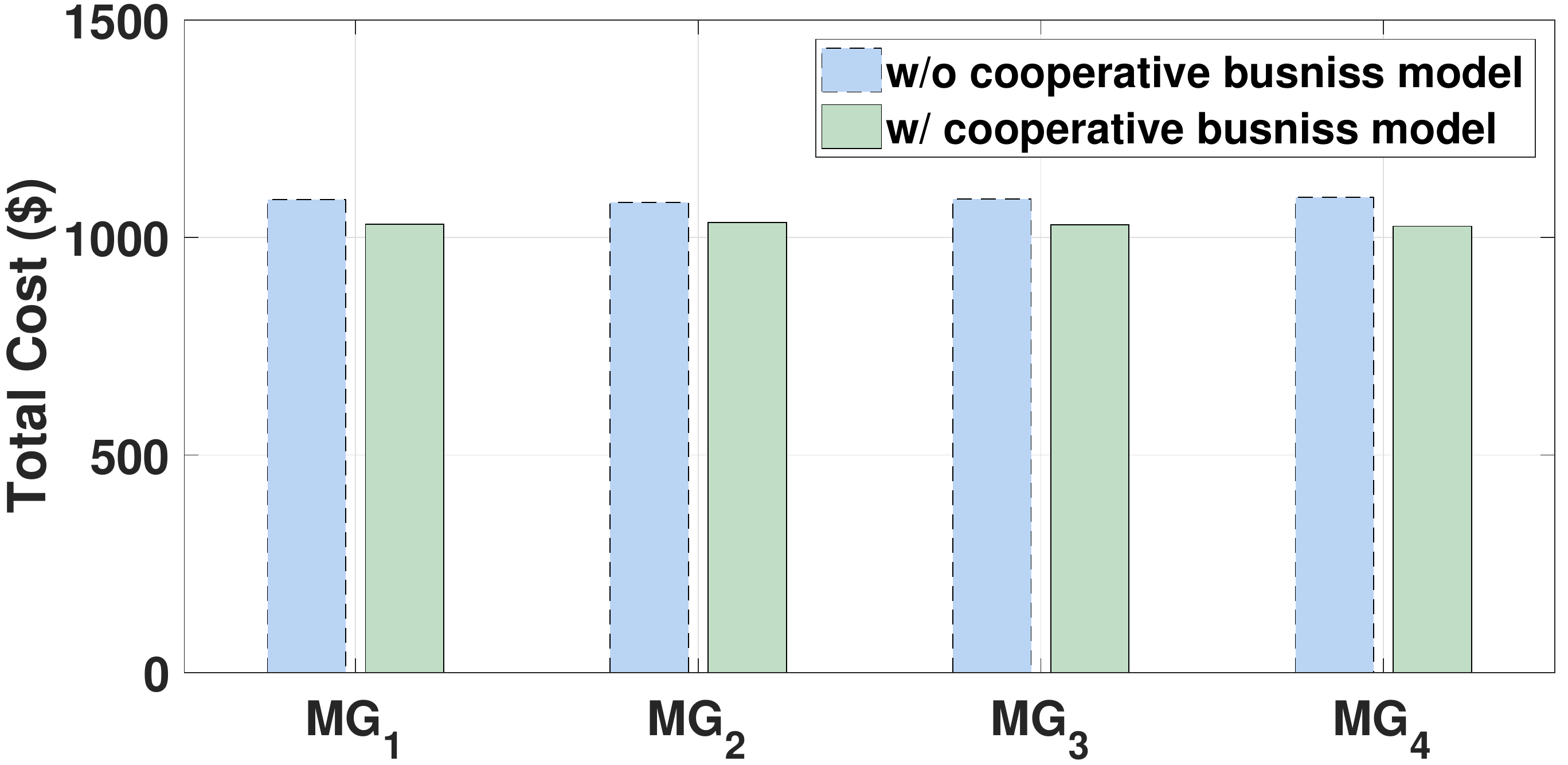}
	\vspace{-5pt}\
	\caption{Comparison of total operational cost of MGs} 
	\centering
	\label{fig.6.4}
		\vspace{-10pt}\
\end{figure}

The locational price signals for the MGs, which are the optimal actions from Level I of the proposed RL-based model, are presented in Fig. \ref{fig.6.2}. Power exchange between MGs and the main grid under optimal price actions, which are the responses of each MG to the actions, are shown in Fig. \ref{fig.6.3}. These figures show the correlation between MGs' behavior and the retail price signal. This demonstrates the mutual impacts of the two levels of the decision model. As the wholesale price increases, the cooperative agent increases the retail prices to encourage the MGs to produce more power to reduce the costs of power purchase from the wholesale market. It can be observed that, most of the time, the cooperative agent exports power to the heavily loaded MGs to maintain power balance in the system. The reason for this is that MGs cannot provide their local demand consumption by their own local generation and have to purchase power from the cooperative service provider. The overall operational costs of MGs have been compared with and without a cooperative agent as an intermediary between the wholesale market and MGs. As can be seen from Fig. \ref{fig.6.4}, the total operational costs of each MG are reduced due to the returned revenue from the cooperative service provider. Therefore, as an intermediary between the MGs and the wholesale market, the cooperative agent can help MGs to reduce their overall operational cost. Hence, it is in the interest of the MGs to participate in the wholesale market through the non-profit cooperative agent.  

\subsection{Benefits of RL-based Method}
A numerical comparison between a centralized off the shelf solver \cite{GAMS} versus the proposed method for the multiple MGs power management problem is shown in Table \ref{table_Com}. In this table, the total social welfare is defined as the summation of the cooperative agent's accumulated reward and the operational cost of all the MGs. Ideally both of the solvers should output the global optimal solution to the problem. As can be seen, the difference between the solutions obtained by the centralized solver with complete system information, and the proposed RL method under incomplete information is less than 0.5\% of the total achieved welfare. Note that while the initial RL training stage can be time-consuming, the decision time is much smaller than that of a centralized optimization method, upon convergence. This is due to the fact that the proposed RL-based method is able to receive continual updates over time, which enables the decision framework to reach a solution in real-time without the need to solve a large-scale optimization problem at each time instant. 
\begin{table}[]
		\centering
		\renewcommand{\arraystretch}{1.3}		
		\caption{Comparison with a Centralized Optimization Method}
		\vspace{-5pt}
		\label{table_Com}
\begin{tabular}{ccc}
\hline
 & RL-based method  & Centralized Opt.\\ \hline
Social welfare (\$)   & 4232.264         & 4212.372     \\ \hline
Computational time (s)   & 9.64         & 116.35     \\ \hline
MG privacy maintenance   & Yes         & No     \\ \hline
\end{tabular}
\end{table}

\begin{figure}
\centering
\subfloat[Optimal action for decision window 1, using the trained models of decision windows 2, 3, and 4 for comparison \label{DW1}]{
\includegraphics[width=0.8\linewidth]{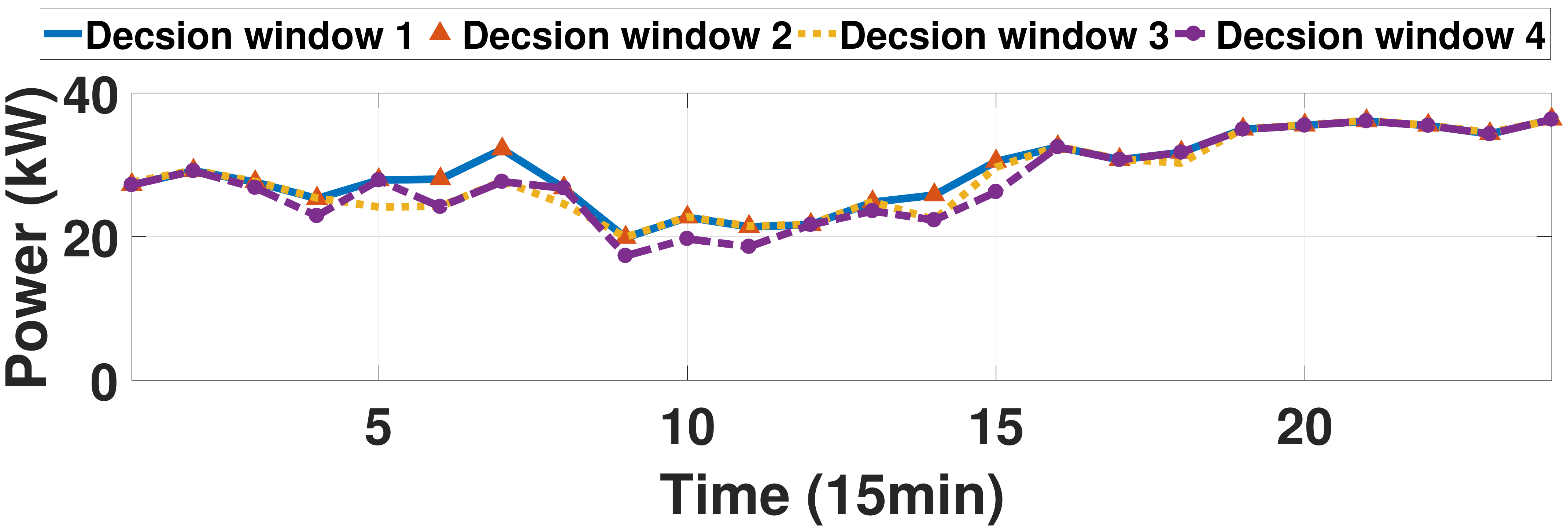}
}
\hfill
\subfloat[Optimal action for decision window 2, using the trained models of decision windows 1, 3, and 4 for comparison \label{DW2}]{
\includegraphics[width=0.8\linewidth]{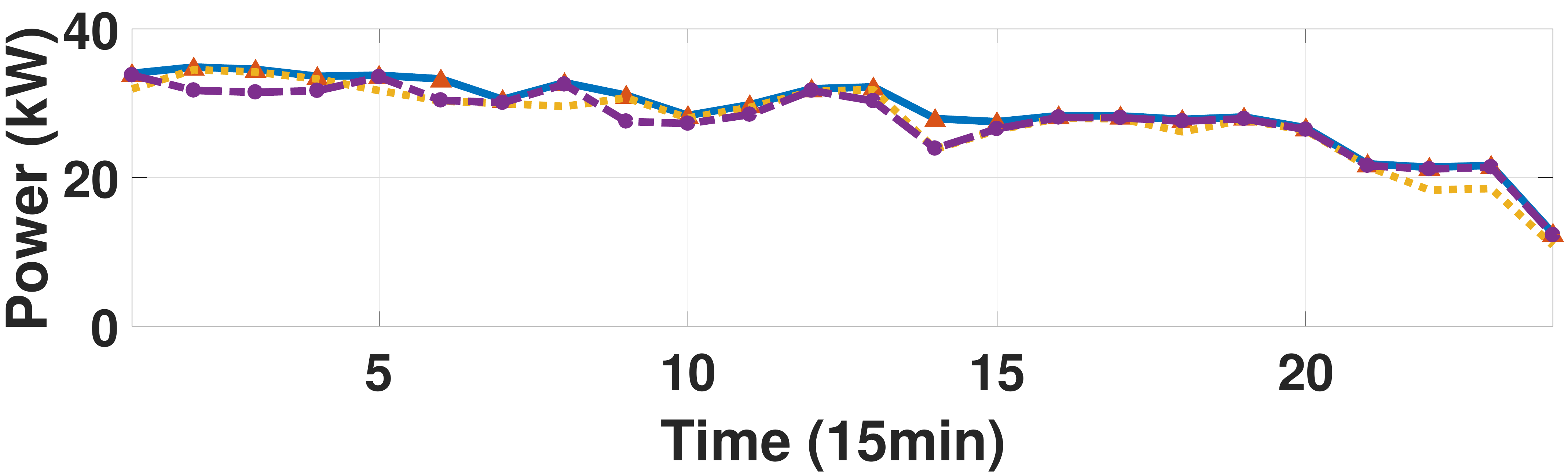}
}
\hfill
\subfloat[Optimal action for decision window 3, using the trained models of decision windows 1, 2, and 4 for comparison\label{DW3}]{
\includegraphics[width=0.8\linewidth]{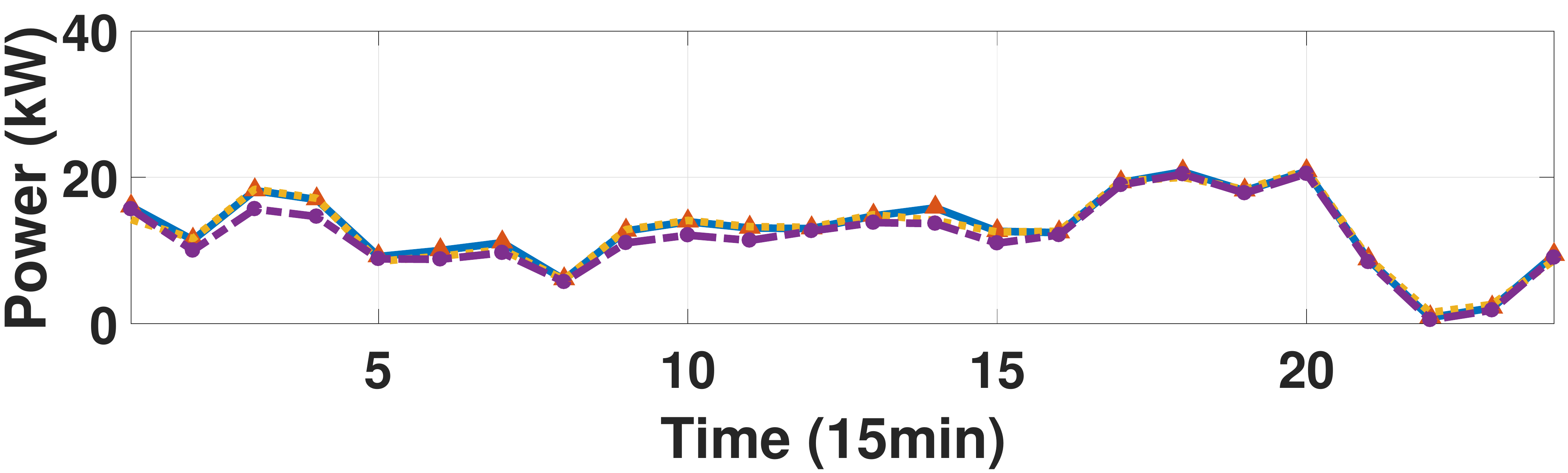}
}
\hfill
\subfloat[Optimal action for decision window 4, using the trained models of decision windows 1, 2, and 3 for comparison\label{DW4}]{
\includegraphics[width=0.8\linewidth]{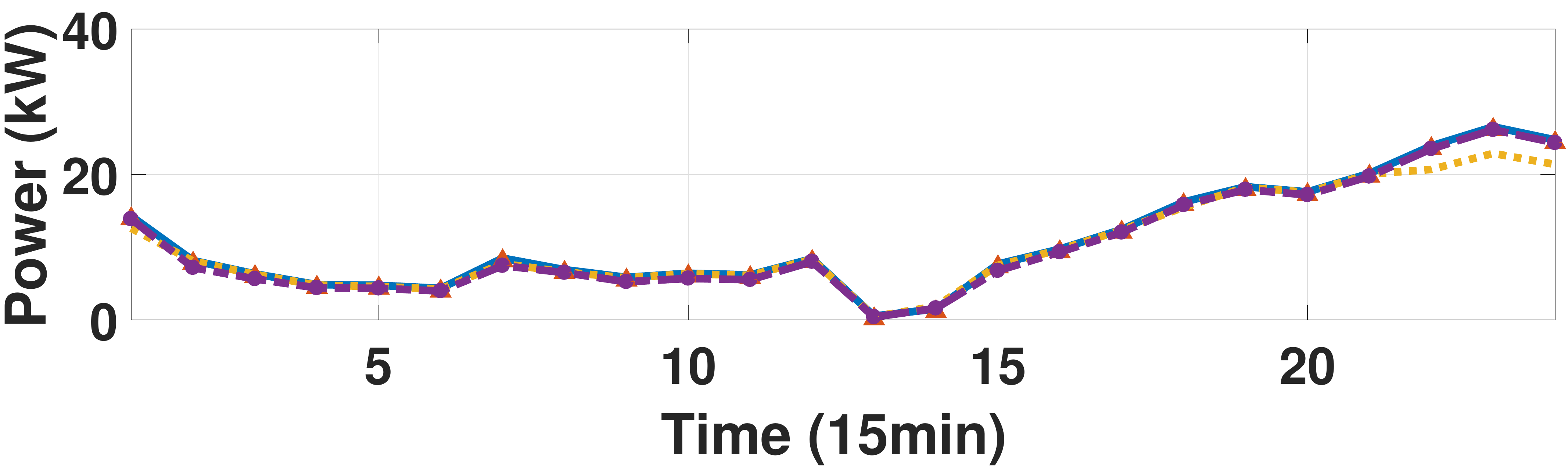}
}
\caption{Verifying the accuracy of previously-learned models under new state scenarios from different decision windows (memory effect).}
\label{fig.coeff}
	\vspace{-5pt} 
\end{figure}

To further demonstrate this, we have performed numerical experiments in which the trained state-action value functions of three different decision windows have been used for a new decision window without re-training. In Fig. \ref{fig.coeff}, optimal power transfers are compared for four scenarios representing four distinct decision windows: in each scenario the RL training is performed for one of the decision windows from random initial conditions, while the updated aggregate MG solar generation and load demand from that decision window are simply inserted into the learned state-action value functions obtained from the other three decision windows. Then, the optimal actions are calculated for each decision window. As can be seen, for all scenarios the optimal solutions are close to each other and almost identical. This shows that the state-action value function learned from other decision windows can be used reliably in new situations using updated state information. Hence, the RL model does not necessarily need to be trained from scratch, and the latest learned function approximator can be simply used to update the cooperative agent's decisions. In practice, however, the re-training process has to be performed with a user-defined frequency depending on the rate of change of system parameters. 

Therefore, the RL-based method has two fundamental advantages over centralized optimization method: 1) RL is model-free; hence, unlike centralized optimization approaches, it does not require detailed private knowledge of MG systems to reach the optimal solution. 2) RL is much faster compared to centralized solvers since the learned state-action value function, which acts similar to a \textit{memory}, is able to leverage the cooperative agent’s past experiences to obtain new optimal solutions by generalizing to new unseen states. 

\subsection{Adaptive RL Results}
\begin{figure}
	\vspace{-0pt} 
	\vspace{-0pt}
	\centering
	\includegraphics[width=0.85\linewidth]{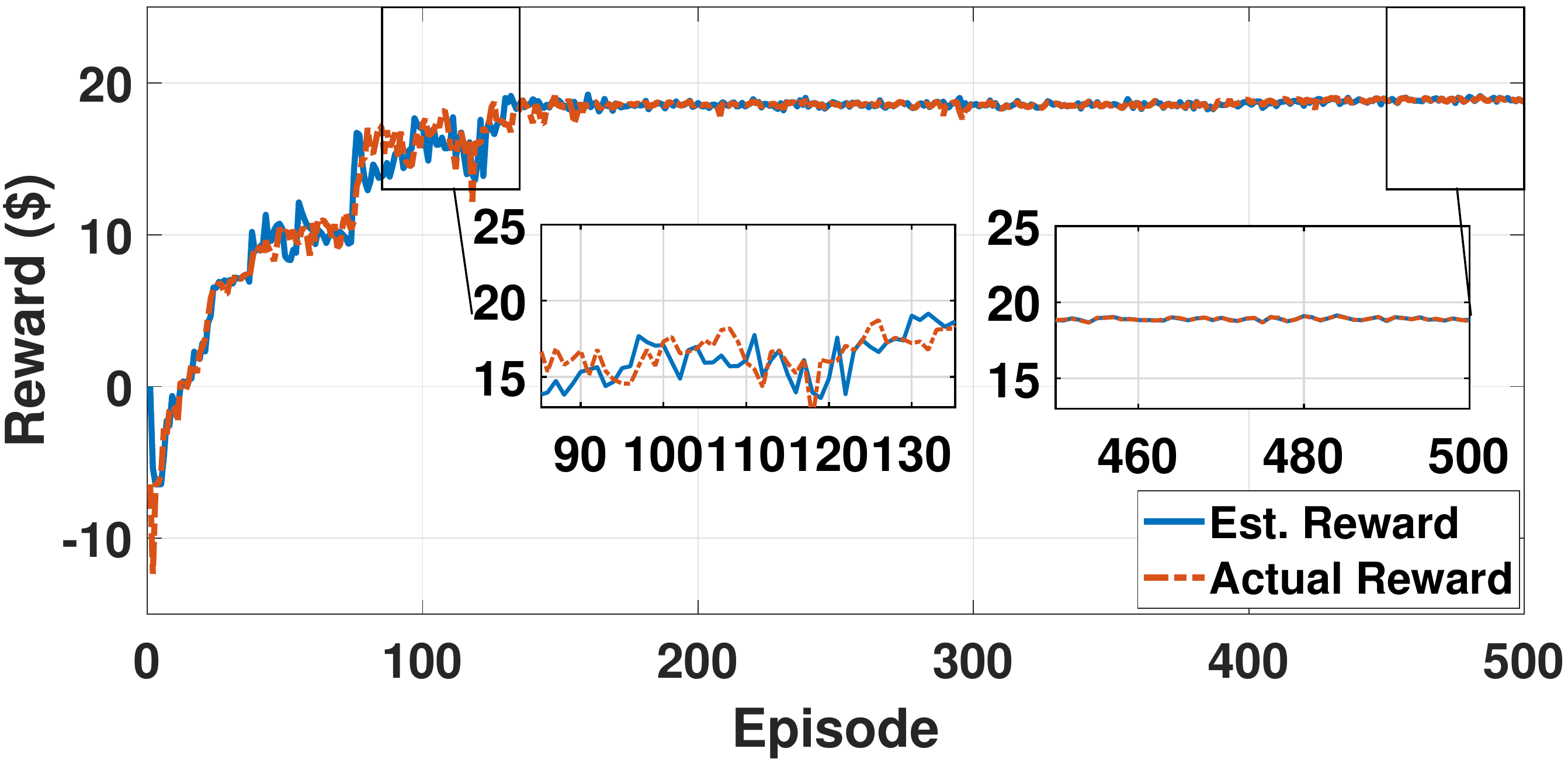}
	\vspace{-5pt}\
	\caption{Performance of the proposed reward function approximation}  
	\centering
	\label{fig.5.6}
		\vspace{-10pt}\
\end{figure}

To verify the functionality of the RL framework, the estimated reward obtained from the multiple linear regression is compared with the actual reward at each episode, as shown in Fig. \ref{fig.5.6}. As can be seen, at the earlier stages of the learning process, the difference between the estimated reward and the real reward is relatively high. However, as the number of episodes increases, this difference drops to within an acceptable range. The results imply that the cooperative agent is able to accurately estimate the response of MGs to control actions. Hence, using the proposed RL approach the cooperative agent is able to track the behavior of MGs and maximize the reward through continuous interactions.  

\begin{figure}
\centering
\subfloat[Estimation MAPE with forgetting factor (Highly adaptive) \label{fig.5.7a}]{
\includegraphics[width=0.85\linewidth]{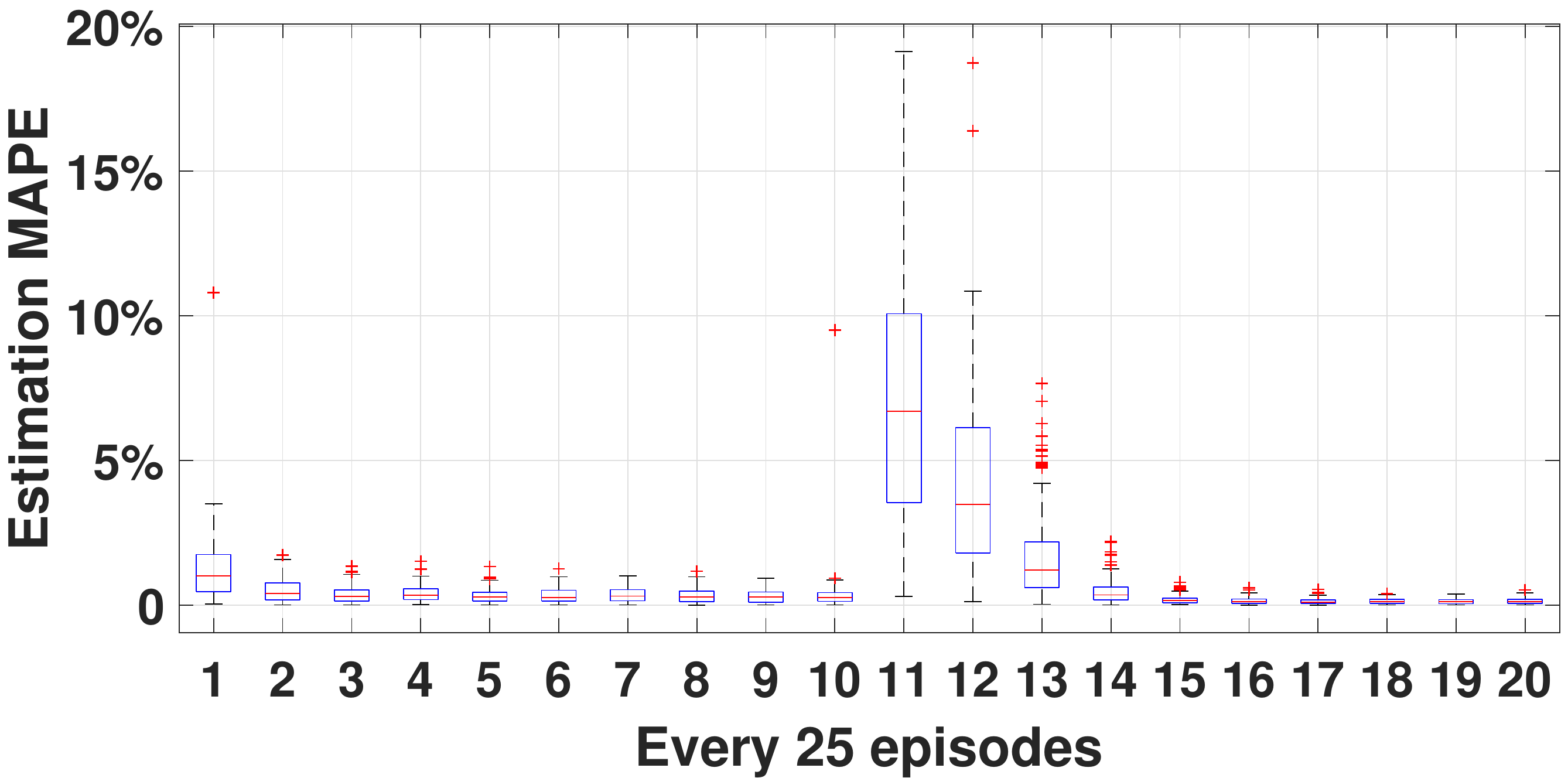}
}
\hfill
\subfloat[Estimation MAPE without forgetting factor (Slow adaptation) \label{fig.5.7b}]{
\includegraphics[width=0.85\linewidth]{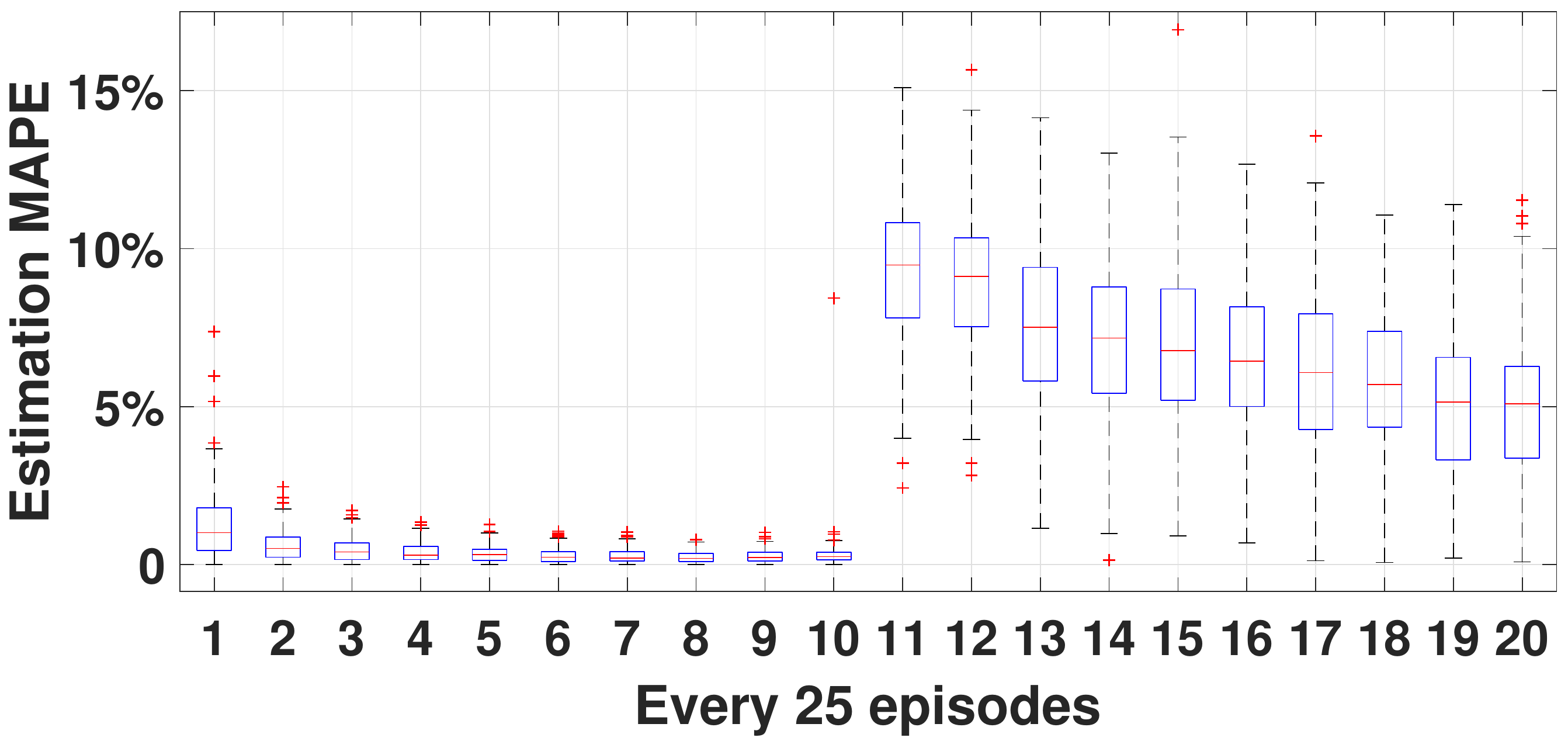}
}
\caption{Adaptability of the proposed RL-based method}
\label{fig.5.7}
		\vspace{-10pt}\
\end{figure}

To test the adaptability of the learning framework against changes in parameters that have not been included in the definition of state set and are not directly observed by the cooperative agent, a numerical scenario is devised. At a point in time (episode $t = 250\ h$), the DG fuel price is doubled. The reward estimation mean absolute percentage error (MAPE) with forgetting factor is shown in Fig. \ref{fig.5.7}(a). As can be seen, upon the occurrence of the sudden change in fuel price, the learning MAPE temporarily jumps to a very high value since the cooperative agent is now facing a new unknown environment, as the price of fuel is not included within the cooperative agent's Markov decision process. However, as the learning process with forgetting proceeds, the MAPE drops to within acceptable range once more. The cooperative agent can still track the actual underlying reward signal as the number of episodes increases with the sudden parameter changes. The reward estimation MAPE without forgetting factor is shown in Fig. \ref{fig.5.7}(b). As can be seen, compared to the proposed adaptive RL-based method with forgetting factor, the conventional RL-based method without forgetting factor shows slow adaptation to changes in parameters. For this case, our RL-method is able to achieve 25\% improvement in the convergence constant over conventional RL.

\begin{figure}
	\vspace{-0pt} 
	\vspace{-0pt}
	\centering
	\includegraphics[width=0.87\linewidth]{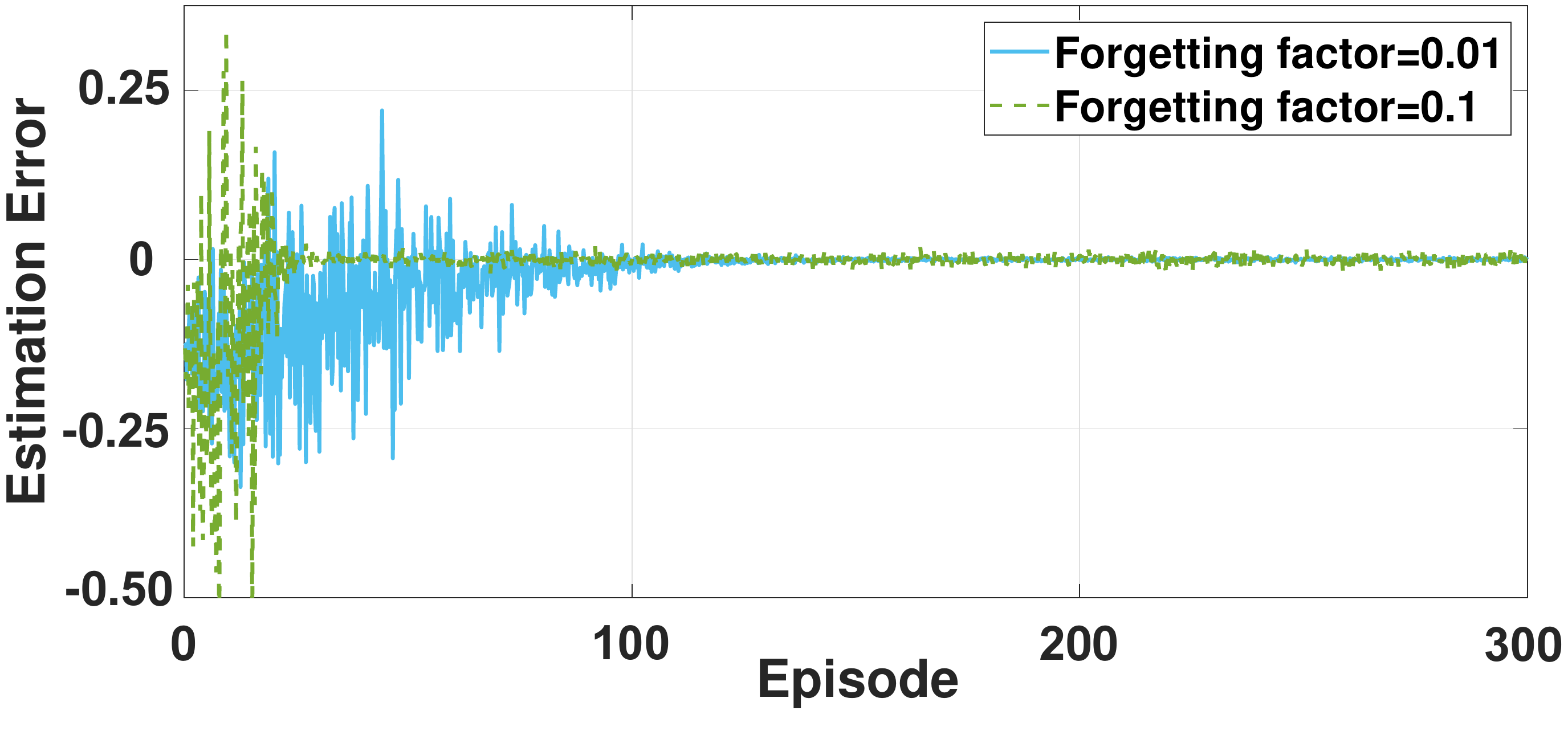}
	\vspace{-5pt}\
	\caption{Impact of forgetting factor on learning convergence}  
	\centering
	\label{fig.5.9}
		\vspace{-10pt}\
\end{figure}

In Fig. \ref{fig.5.9}, the impact of forgetting factor on the convergence of the RL framework is demonstrated. This figure shows the RL-based reward estimation error for the cooperative agent under two different forgetting factor values. As the forgetting factor increases from 0.01 to 0.1, the convergence speed of the RL framework has been improved. Hence, the forgetting factor controls the rate of adaptiveness to new conditions. However, a tradeoff exists between the rate of convergence and the accuracy of the solution. As can be seen, higher forgetting factors also lead to higher variances in the estimation error signal.

\section{Conclusions}\label{sec:Con}
Smart distribution systems with networked MGs in a cooperative setting can facilitate reliable power delivery to customers in future rural power grids. However, cooperatives can have incomplete knowledge of MG members' operational parameters due to data privacy and ownership concerns, which is an obstacle in the way of optimal decision making. Motivated by the shortcomings of model-based multiple MG power management in distribution systems with limited observability, this paper presents an adaptive RL-based methodology for bi-level power management of cooperatives consisting of multiple networked MGs.

We have shown that: 1) using the proposed decision method, a cooperative agent is able to accurately track the behavior of multiple networked MGs under incomplete knowledge of operation variables behind the PCCs. This can be used to indirectly control the response of participants in a price-based environment. 2) The proposed RL-based method is able to generalize from its past experiences to estimate optimal solutions in new situations without re-training from random initial conditions (i.e., fast response under evolving system conditions). This immensely speeds up the power management computational process. 3) The framework is shown to be adaptive against the changes happening to unobserved parameters that are excluded from cooperative agent's state set. The learning model has been tested and verified using extensive numerical scenarios. To summarize, the proposed decision model shows better adaptability, solution quality, and computational time compared to conventional centralized optimization methods. 

The current RL-based decision model is limited to the power management of a single cooperative service provider with multiple MGs. However, in more realistic cases, there could also be multiple cooperative service providers in an interconnected rural area, which implies that the impact of cooperative service providers on each other and on the wholesale price could not be ignored. Hence, an optimal coordination scheme needs to be designed to enable collaboration among multiple entities. In future work, we will extend the proposed RL method to address this challenge.  
 
\appendix[MG Optimal Power Management Formulation]
A moving look-ahead decision window $[t,t+T]$ is defined using the latest estimations of solar and load power at different instants, where $n$ is the MG index ($n\in\{1,...,N\}$), $i$ and $j$ define the bus numbers for each MG ($\forall i,j\in \Omega_I$), and $k$ denotes the line index ($\forall k\in \Omega_K$). It has decision vector $\pmb{x_p} = (P^{DG}_{i,t,n},P^{PCC}_{t,n},P^{Ch}_{i,t,n},P^{Dis}_{i,t,n})^\top$ and $\pmb{x_q} = (Q^{DG}_{i,t,n},Q^{PCC}_{t,n},Q^{PV}_{i,t,n},Q^{ESS}_{i,t,n})^\top$. 
\begin{equation}
\min_{\pmb{x_p},\pmb{x_q}}\ \sum_{t}^{T+t}(-\lambda^{R}_{t,n}P^{PCC}_{t,n}+\lambda^{F}_{i,t,n}F_{i,t,n})\label{eq3_1}
\end{equation}
\begin{equation}
s.t.\ \ \ F_{i,t,n}=a_{f}(P^{DG}_{i,t,n})^2+b_{f}P^{DG}_{i,t,n}+c_{f}\label{eq3_2}
\end{equation}
\begin{equation}
|P^{PCC}_{t,n}|\leq P_{t,n}^{PCC,M}\label{eqPCC}
\end{equation}
\begin{equation}
|Q^{PCC}_{t,n}|\leq Q_{t,n}^{PCC,M}\label{eqPCCQ}
\end{equation}
\begin{equation}
0\leq P^{DG}_{i,t,n}\leq P_{i,n}^{DG,M}\label{eq3_3}
\end{equation}
\begin{equation}
0\leq Q^{DG}_{i,t,n}\leq Q_{i,n}^{DG,M}\label{eq3_3q}
\end{equation}
\begin{equation}
|P^{DG}_{i,t,n}-P^{DG}_{i,t-1,n}|\leq P_{i,n}^{DG,R}\label{eq3_4}
\end{equation}
\begin{equation}
P^{ij}_{t,n}=V^{i}_{t,n}(V^{i}_{t,n}G_{n}^{ij}-V^{j}_{t,n}(G_{n}^{ij}cos(\Delta\theta^{ij}_{t,n})+B_{n}^{ij}sin(\Delta\theta^{ij}_{t,n})))\label{eq3_51}
\end{equation}
\begin{equation}
Q^{ij}_{t,n} = -V^{i}_{t,n}(V^{i}_{t,n}B_{n}^{ij}+V^{j}_{t,n}(G_{n}^{ij}cos(\Delta\theta^{ij}_{t,n})-B_{n}^{ij}sin(\Delta\theta^{ij}_{t,n})))\label{eq3_52}
\end{equation}
\begin{equation}
(P^{ij}_{t,n})^2 +  (Q^{ij}_{t,n})^2\leq (L^{ij,M}_{t,n})^2 \label{eq3_50}
\end{equation}
\begin{equation}
\sum_{ij \in k}^{K} P_{t,n}^{ij}= \sum_{ji \in k}^{K} P_{t,n}^{ji}-p_{i,t,n}\label{eq3_5}
\end{equation}
\begin{equation}
\sum_{i,j \in k}^{K} Q_{t,n}^{ij}= \sum_{j,i \in k}^{K} Q_{t,n}^{ji}-q_{i,t,n}\label{eq3_6}
\end{equation}
\begin{equation}
p_{i,t,n}= P_{i,t,n}^{D,e}-P_{i,t,n}^{DG}-P_{i,t,n}^{PV,e}+P^{Ch}_{i,t,n}-P^{Dis}_{i,t,n}\label{eq3_5_1}
\end{equation}
\begin{equation}
P_{i,t,n}^{D}=P_{i,t,n}^{D,e}-\varepsilon^{D}_{i,t,n}\label{eq3_5_2}
\end{equation}
\begin{equation}
P_{i,t,n}^{PV}=P_{i,t,n}^{PV,e}-\varepsilon^{PV}_{i,t,n}\label{eq3_5_3}
\end{equation}
\begin{equation}
q_{i,t,n}= Q_{i,t,n}^{D}-Q_{i,t,n}^{DG}-Q_{i,t,n}^{PV}+Q^{ESS}_{i,t,n}\label{eq3_6_1}
\end{equation}
\begin{equation}
V^{PCC}_{t,n}=V_{t,n}^{PCC,E}\label{eq3_8}
\end{equation}
\begin{equation}
V_{i,n}^{m}\leq V_{i,t,n}\leq V_{i,n}^{M}\label{eq3_9}
\end{equation}
\begin{equation}
|Q_{i,t,n}^{PV}|\leq Q_{i,n}^{PV,M}\label{eq3_pv}
\end{equation}
\begin{equation}
SOC_{i,t,n} = SOC_{i,t-1,n}+\Delta t (P^{Ch}_{i,t,n}\eta_{Ch}- P^{Dis}_{i,t,n}/\eta_{Dis})/E^{Cap}_{i,n}\label{eq3_10}
\end{equation}
\begin{equation}
SOC_{i,n}^{m}\leq SOC_{i,t,n}\leq SOC_{i,n}^{M}\label{eq3_11}
\end{equation}
\begin{equation}
0\leq P^{Ch}_{i,t,n}\leq u^{Ch}_{i,t,n}P^{Ch,M}_{i,n}\label{eq3_12}
\end{equation}
\begin{equation}
0\leq P^{Dis}_{i,t,n}\leq u^{Dis}_{i,t,n}P^{Dis,M}_{i,n}\label{eq3_13}
\end{equation}
\begin{equation}
0\leq u^{Ch}_{i,t,n}+u^{Dis}_{i,t,n}\leq 1\label{eq3_14}
\end{equation}
\begin{equation}
u^{Ch}_{i,t,n},u^{Dis}_{i,t,n}\in \{0, 1\}\label{eq3_15}
\end{equation}

The objective function \eqref{eq3_1} minimizes each MG's total cost of operation, which is composed of two terms: the negative of revenue from power transfer with the cooperative agent and the cost of running local DGs. Here, $\lambda^{F}_{t,n}$ is the diesel generator fuel price in $\$/L$ adopted from \cite{EIA_fuel}. The fuel consumption $F_{i,t,n}$ of diesel generator can be expressed as a quadratic polynomial function \eqref{eq3_2}, with coefficients $a_{f}=0.0001773\ L/kW^2$, $b_{f}=0.1709\ L/kW$, and $c_{f}=14.67 L$ adopted from \cite{Ali2015}. Constraints \eqref{eqPCC}-\eqref{eqPCCQ} describe the power exchange limit between the MG and the upstream distribution grid with the maximum active/reactive power exchange limits, $P_{t,n}^{PCC,M},Q_{t,n}^{PCC,M}$. Constraints \eqref{eq3_3}-\eqref{eq3_3q} ensure that the DG active/reactive power outputs, $P^{DG}_{i,t,n}/Q^{DG}_{i,t,n}$, are within the DG power capacity $P_{i,n}^{DG,M},Q_{i,n}^{DG,M}$, and \eqref{eq3_4} enforces the maximum DG ramp limit, $P_{i,n}^{DG,R}$. Internal AC power flow model of the MG is considered here with the network topology constraints, with \eqref{eq3_51} and \eqref{eq3_52} determining the active and reactive power flows of each branch, where $G^{ij}$ and $B^{ij}$ are the corresponding real and imaginary parts of the bus admittance matrix, and $V^{i}_{t,n}$ and $\Delta\theta^{ij}_{t,n}$ are the nodal voltage magnitude and phase angle difference, respectively. Constraint \eqref{eq3_50} denotes the power flow limits for each branch. Equations \eqref{eq3_5}-\eqref{eq3_6_1} are the nodal active/reactive power balances at MG buses. The difference between the predicted and actual PV/load values are modeled using Gaussian error variables as shown in equations \eqref{eq3_5_2} and \eqref{eq3_5_3}, where $P_{i,t,n}^{D,e}$ denotes the estimated active load, and $P_{i,t,n}^{PV,e}$ is the estimated active power output of PV. Also, $\varepsilon^{D}_{i,t,n},\varepsilon^{PV}_{i,t,n}\sim N(0,\sigma)$ denote the Gaussian estimation errors for active load and PV power, respectively. Constraint \eqref{eq3_8} sets the voltage at the PCC of the MG according to the estimated input voltage, $V_{t,n}^{PCC,E}$. Constraint \eqref{eq3_9} sets the limits for nodal bus voltage amplitude, $[V_{i,n}^{m},V_{i,n}^{M}]$. PV reactive power output, $Q_{i,t,n}^{PV}$, is constrained by its maximum limit $Q_{i,n}^{PV,M}$ in \eqref{eq3_pv}. Operational ESS constraints are described by \eqref{eq3_10}-\eqref{eq3_15}. Constraint \eqref{eq3_10} determines the state of charge (SOC) of ESSs, $SOC_{i,t,n}$. The SOC and charging/discharging power of ESS, $P^{Ch}_{i,t,n}$, $P^{Dis}_{i,t,n}$, are constrained in \eqref{eq3_11}-\eqref{eq3_15}. Here, $[SOC_{i,n}^{m},SOC_{i,n}^{M}]$, $P^{Ch,M}_{i,n}$ and $P^{Dis,M}_{i,n}$ define the permissible range of SOC, and maximum charging and discharging power, with $u^{Ch}_{i,t,n}$ and $u^{Dis}_{i,t,n}$ denoting the charge/discharge binary indicator variables, and $\eta_{Ch}$/$\eta_{Dis}$ representing the charging/discharging efficiency. $E^{Cap}_{i,n}$ denotes the maximum capacity of ESSs. 






\ifCLASSOPTIONcaptionsoff
  \newpage
\fi



\bibliographystyle{IEEEtran}
\bibliography{IEEEabrv,./bibtex/bib/IEEEexample}
\end{document}